# Utility-based Link Recommendation for Online Social Networks

Zhepeng Li*
Schulich School of Business
York University
zli@ schulich.yorku.ca

Xiao Fang*
Lerner College of Business and Economics
University of Delaware
xfang@udel.edu

Xue Bai
School of Business
University of Connecticut
xue.bai@ uconn.edu

Olivia Sheng
David Eccles School of Business
University of Utah
olivia.sheng@eccles.utah.edu

*Equal Contribution. Send comments to xfang@udel.edu.

# Utility-based Link Recommendation for Online Social Networks

## ABSTRACT

Link recommendation, which suggests links to connect currently unlinked users, is a key functionality offered by major online social networks. Salient examples of link recommendation include "People You May Know" on Facebook and LinkedIn as well as "You May Know" on Google+. The main stakeholders of an online social network include users (e.g., Facebook users) who use the network to socialize with other users and an operator (e.g., Facebook Inc.) that establishes and operates the network for its own benefit (e.g., revenue). Existing link recommendation methods recommend links that are likely to be established by users but overlook the benefit a recommended link could bring to an operator. To address this gap, we define the utility of recommending a link and formulate a new research problem – the utility-based link recommendation problem. We then propose a novel utility-based link recommendation method that recommends links based on the value, cost, and linkage likelihood of a link, in contrast to existing link recommendation methods which focus solely on linkage likelihood. Specifically, our method models the dependency relationship between value, cost, linkage likelihood and utility-based link recommendation decision using a Bayesian network, predicts the probability of recommending a link with the Bayesian network, and recommends links with the highest probabilities. Using data obtained from a major U.S. online social network, we demonstrate significant performance improvement achieved by our method compared to prevalent link recommendation methods from representative prior research.

## 1. Introduction

Online social networks, such as Facebook, LinkedIn, and Google+, have gained unprecedented numbers of users in a short time period, attracting massive attention from both industry and academia to study and utilize these networks for economic and societal benefits (Jackson 2008, Backstrom and Leskovec 2011, Fang et al. 2013a). It is common for online social networks to implement a link recommendation mechanism, which suggests links to connect currently unlinked users. As shown in Figure 1, salient examples of link recommendation include “People You May Know” on Facebook and LinkedIn as well as “You May Know” on Google+. Since its early success on LinkedIn, link recommendation has become a standard feature of online social networks (Davenport and Patil 2012).

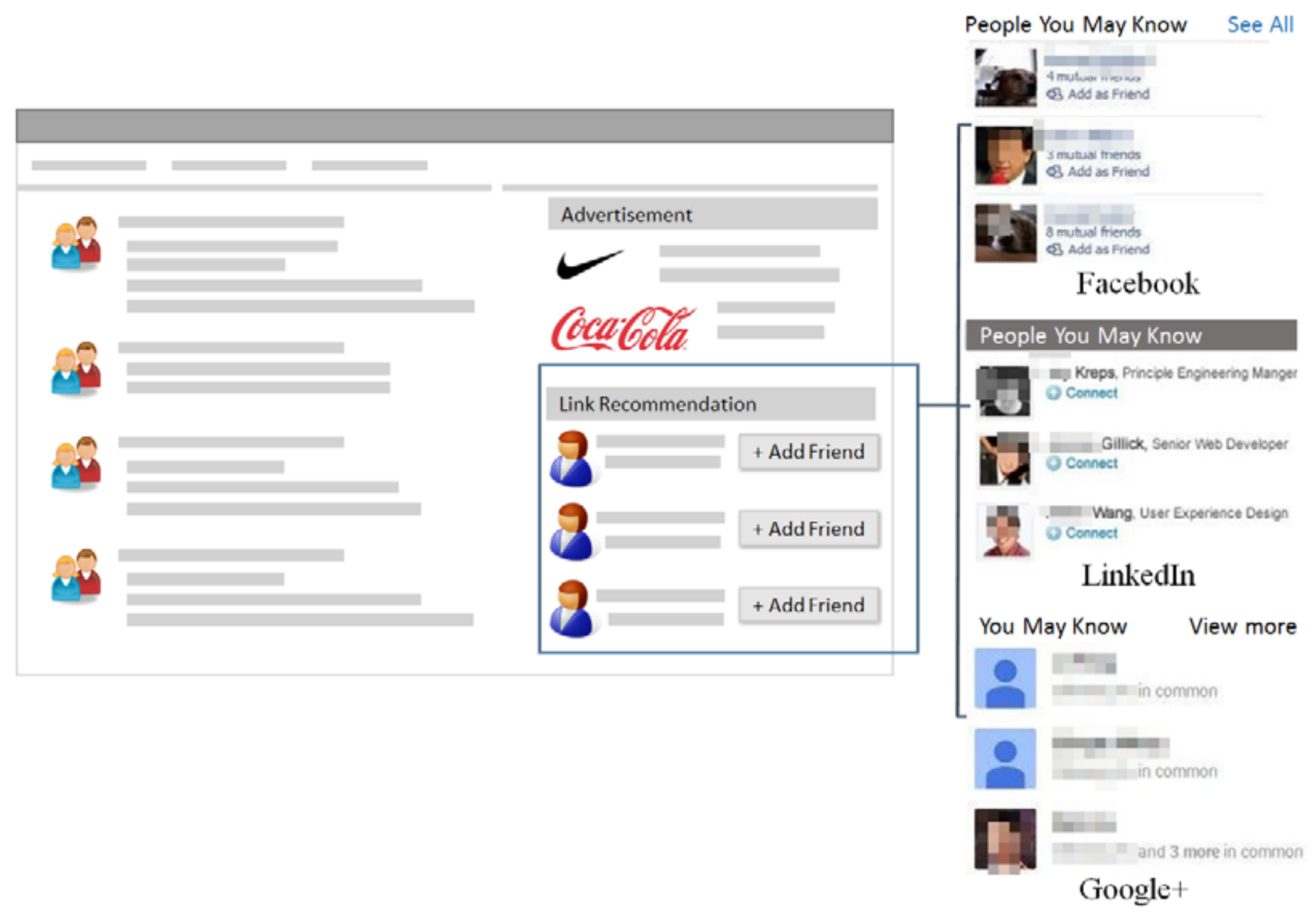

**Figure 1: Link Recommendation in Online Social Networks**

The main stakeholders of an online social network include users (e.g., Facebook users) who use the network to connect and communicate with other users (i.e., friends) and an operator (e.g., Facebook Inc.) that establishes and operates the network for its own benefit (e.g., revenue) (Ellison et al. 2007, Huberman et al. 2009, Kaplan and Haenlein 2010). Therefore, the advantages of link recommendation are twofold. First, link recommendation could cater to users’ needs of socializing and networking with others in an online social network. By helping users connect with new friends, link recommendation allows new users to quickly become engaged in a community and facilitates existing users to enlarge their circles of friends. Second, link recommendation could benefit the operator of an online social network as well. According to eMarketer.com (2012), operators of online social networks reaped an estimated $12 billion from advertisements on these networks in 2014, up from $10 billion in 2013. Understandably, link recommendation potentially leads to a more connected network of users, which drives advertisements to reach more users and ultimately brings more revenue to the network's operator. Existing research develops link recommendation methods from the perspective of link prediction (Hasan et al. 2006, Liben-Nowell and Kleinberg 2007, Lichtenwalter et al. 2010, Gong et al. 2012). In general, these methods predict the likelihood that a potential link[1] will be established by users, namely linkage likelihood, and recommend potential links with the highest linkage likelihoods (Hasan et al. 2006, Liben-Nowell and Kleinberg 2007, Lichtenwalter et al. 2010, Gong et al. 2012). While existing link recommendation methods cater well to users' social needs by recommending links that are likely to be established, they largely overlook the other advantage of link recommendation, i.e., benefiting the operator of an online social network; this is a fundamental gap that motivates our research.

We illustrate the gap using the example of Facebook, whose operator harvests the majority of its $7.9 billion revenue from advertisements on the network (Facebook 10-K 2013). Facebook allows an advertisement to be placed on the Facebook page of selected users. A user can interact with the advertisement through actions including click, comment, like, and share. Such interaction propagates the advertisement to the user's friends, who can also interact with the advertisement and further propagate it to their friends. As this propagation process continues, the advertisement can reach a much larger number of users than the initially selected users. Facebook Inc. obtains revenue each time the advertisement reaches a user. In this context, let us consider recommending one link out of the two potential links $e_{15}$ and $e_{16}$, shown in Figure 2. Assuming that $e_{15}$ and $e_{16}$ have the same linkage

[1] A potential link refers to a link that has not been established.

likelihood, existing methods that recommend links purely based on linkage likelihood are indifferent about them and randomly pick one to recommend (Hasan et al. 2006, Liben-Nowell and Kleinberg 2007, Lichtenwalter et al. 2010, Gong et al. 2012). However, $e_{15}$ could bring much more advertisement revenue to Facebook Inc. than $e_{16}$ because of the following considerations. First, advertisements initially placed on the Facebook page of user $w_1$ could reach more users through $e_{15}$ than $e_{16}$. Second, advertisements propagated from users $w_2$, $w_3$, and $w_4$ to user $w_1$ could reach more users through $e_{15}$ than $e_{16}$. Third, more users could propagate advertisements to users $w_1$, $w_2$, $w_3$, and $w_4$ through $e_{15}$ than $e_{16}$. Therefore, for the benefit of Facebook Inc., it is much more desirable to recommend $e_{15}$ than $e_{16}$, while existing link recommendation methods are indifferent about them.

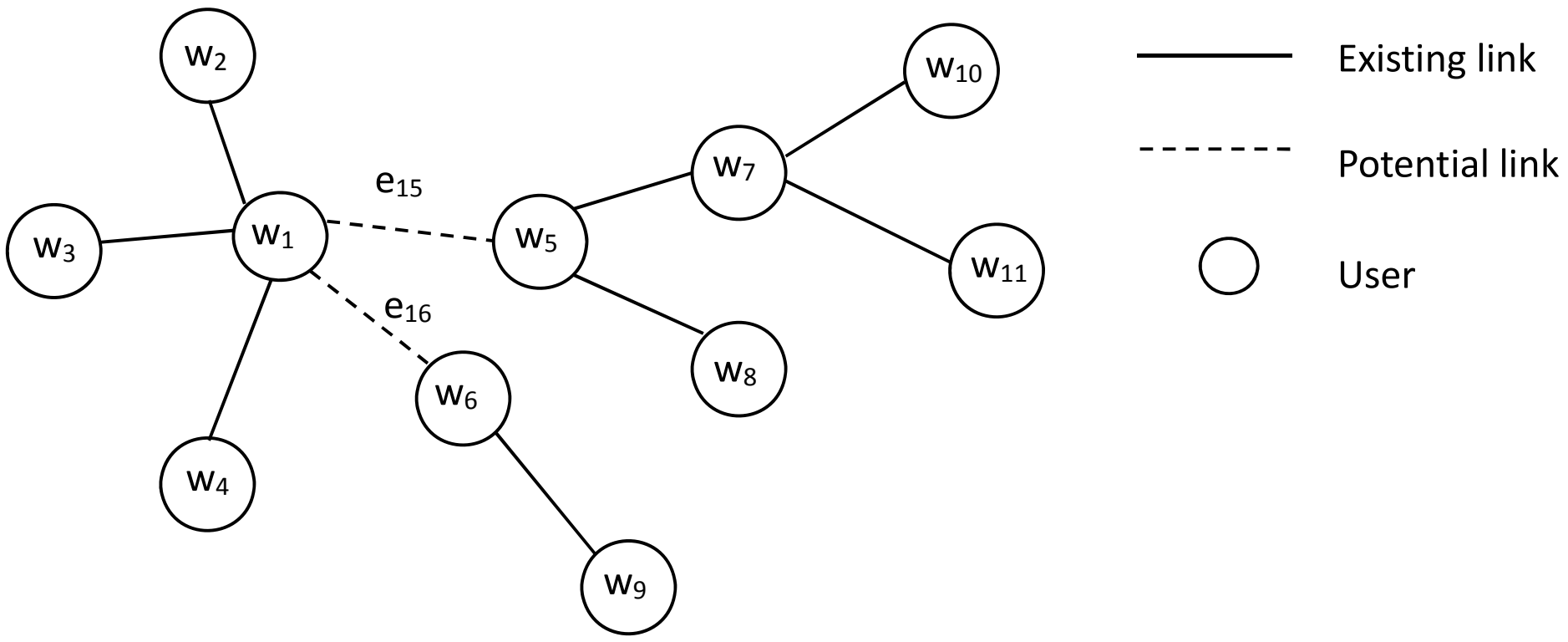

**Figure 2:** An Illustrating Example

To address this gap, this study defines a new link recommendation problem and proposes a novel link recommendation method. The key difference between our link recommendation problem and link recommendation problems defined in prior studies is the consideration of the operator's benefit from link recommendation in our problem formulation. Since different potential links occupy different structural positions in an online social network, they could each bring different values to the network's operator. Furthermore, the value of a potential link can only be realized if it is established by users; on the other hand, a cost to the network's operator is incurred if a recommended link is not established. Therefore, we define the utility of recommending a potential link by considering its value, cost, and whether it will be established, and formulate the utility-based link recommendation problem. To solve the problem, we propose a novel utility-based link recommendation method. Unlike existing link recommendation methods that recommend potential links solely based on their linkage likelihoods, our method considers their values, costs, and linkage likelihoods when making link recommendation decisions. Specifically, we propose a Bayesian network learning method that models the dependency relationship between value, cost, linkage likelihood and link recommendation decision, predicts the probability of recommending a potential link with the learned Bayesian network, and recommends potential links with the highest probabilities. We note that linkage likelihood is latent (unobserved) and continuous. Hence, the principal methodological obstacle overcome by our proposed method is how to learn a Bayesian network with a continuous latent factor.

The rest of the paper is organized as follows. We start in §2 by reviewing prior works related to our study. We then define the utility-based link recommendation problem in §3 and propose a utility-based link recommendation method in §4. The effectiveness of our proposed method is evaluated using data collected from a major online social network in §5. The paper concludes with implications and future research directions in §6.

## 2. Related Work

Link recommendation methods proposed in prior studies predict the linkage likelihood that a potential link will be established and recommend potential links with the highest linkage likelihoods. According to different prediction approaches used, prior link recommendation methods can be broadly categorized into learning-based link recommendation methods and proximity-based link recommendation methods. In the following, we review representative methods in each category.

Learning-based link recommendation methods learn a model from observed link establishments and predict linkage likelihood using the learned model. Given a social network, one can construct training data from observed link establishments in the network. Generally, each record of the training data has the format $< \mathrm{f}_1, \mathrm{f}_2, \ldots, \mathrm{f}_\mathrm{m}, \mathrm{c} >$, where $\mathrm{f}_1, \mathrm{f}_2, \ldots, \mathrm{f}_\mathrm{m}$ represent features that affect link establishment and c is the class label. The class label c is 1 for an existing link and it is 0 for a potential link. Commonly used features include topological features that are derived from the structure of a social network and nodal features that are computed from the characteristics of individual users in a social network. Topological features such as the number of common neighbors and the shortest distance between users have frequently been used by learning-based link recommendation methods (O'Madadhain et al. 2005, Hasan et al. 2006, Wang et al. 2007, Benchettara et al. 2010, Lichtenwalter et al. 2010). Nodal features that are calculated from users' demographical and geographical characteristics have also been widely employed (O'Madadhain et al. 2005, Zheleva et al. 2008, Scellato et al. 2011, Wang et al. 2011).

Once training data are constructed, supervised machine learning methods can be applied to the data to predict linkage likelihood. O'Madadhain et al. (2005) employ logistic regression to predict the likelihood of interaction between users using data collected from CiteSeer and Enron emails. Wang et al. (2007) combine a local Markov random field model and logistic regression to predict the likelihood of co-authorship using DBLP and PubMed article datasets. Benchettara et al. (2010) also target predicting the likelihood of co-authorship but employ a decision tree classifier enhanced with Adaboost for this prediction. Hopcroft et al. (2011) adopt a factor-based graphical model to predict reciprocal relationships on Twitter. Gong et al. (2012) predict linkage likelihood in Google+ using support vector machine (SVM). Besides classification methods, other supervised learning methods such as supervised random walk (Backstrom and Leskovec 2011), matrix factorization-based methods (Kunegis et al. 2010, Yang et al. 2011), and relational learning (Popescul and Ungar 2003) have been employed to predict linkage likelihood in unipartite or bipartite social networks.

Proximity-based link recommendation methods surrogate the linkage likelihood of a potential link using the proximity between users that would be connected by the link. According to McPherson et al. (2001), similar users are more likely to interact and connect with each other. Therefore, higher proximity indicates higher chance of linkage. Proximity metrics employed by link recommendation methods consist of nodal proximity metrics and structural proximity metrics. Nodal proximity metrics measure the similarity between users using their characteristics, including demographical characteristics, such as age, education, and occupation (Zheleva et al. 2010), geographical characteristics, such as co-location and distance (Quercia and Capra 2009, Crandall et al. 2010, Wang et al. 2011), and semantic characteristics, such as keywords and annotation tags (Shen et al. 2006, Chen et al. 2009, Schifanella et al. 2010, Kuo et al. 2013). To compute nodal proximity between users, typical similarity functions such as the Manhattan distance, the cosine similarity, the KL-divergence, and the Jaccard coefficient, have been applied to users' characteristics (Shen et al. 2006, Chen et al. 2009, Scellato et al. 2011, Wang et al. 2011, Adali et al. 2012).

Users' structural features have been widely employed to study their behaviors in social networks. For example, Doreian (1989) and Zhang et al. (2013) propose autocorrelation models to examine the impact of users' structural features such as cohesion and structural equivalence on their actions and choices in social networks. In link recommendation, structural proximity metrics measure the proximity between users using their structural features in a social network (Liben-Nowell and Kleinberg 2007). One type of structural proximity metrics targets users' neighborhoods. For example, the common neighbor between users is defined as the number of their mutual neighbors in a social network (Newman 2001, Liben-Nowell and Kleinberg 2007). Extended from the common neighbor metric, the Adamic/Adar metric assigns less weight for more connected common neighbors (Adamic and Adar 2003, Liben-Nowell and Kleinberg 2007). Motivated by the finding that the likelihood of linking two users is correlated with their neighborhood sizes, the preferential attachment between users is defined as the product of their neighborhood sizes (Barabási and Albert 1999, Newman 2001, Barabási et al. 2002, Liben-Nowell and Kleinberg 2007). Observing that two users are similar if their neighbors are similar, Jeh and Widom (2002) define the SimRank score between users as the average of their neighbors' SimRank scores. Going beyond neighborhoods, another type of structural proximity metrics focuses on the paths connecting users. The Katz index measures the structural proximity between users using the number of paths connecting them, weighted by the lengths of these paths (Katz 1953). Originally developed for measuring the social status of a social entity, the Katz index has been shown to be effective in forecasting linkage likelihood (Liben-Nowell and Kleinberg 2007). Considering link establishment between users as a random walk from one to the other, Tong et al. (2006) adapts the PageRank algorithm (Brin and Page 1998) to compute the structural proximity between users as the summation of stationary probabilities that one user reaches the other. Based on a similar idea of treating link establishment as a random walk, Fouss et al. (2007) define the hitting time between users as the expected number of steps to reach one user from the other for the first time and measure the structural proximity between them as the negation of their hitting time.

Our literature review suggests that existing link recommendation methods focus on predicting linkage likelihood but overlook the benefit of link recommendation to an operator. To overcome this limitation, we define a new link recommendation problem that takes into account operator's benefit from link recommendation. We then propose a novel link recommendation method to solve the problem. The essential novelty of our proposed method lies in its consideration of the value, cost and linkage likelihood of a potential link when making a link recommendation decision, in contrast to existing methods that recommend links based solely on linkage likelihood.

## 3. Utility-based Link Recommendation Problem

In this section, we define the utility of recommending a potential link and formulate the utility-based link recommendation problem. Let $W = \{w_j\},\ j = 1, 2, \dots, n,$ be a set of users in an online social network. User $w_j$ makes value contribution $v_j$ to the operator of the online social network (Granovetter 2005, Jackson 2008). Value $v_j$ consists of two parts: intrinsic value and network value, and

$$v_j = v_j^I + v_j^N, \quad (1)$$

where $v_j^I$ and $v_j^N$ represent the intrinsic and network value of $w_j$ respectively (Jackson and Wolinsky 1996, Domingos and Richardson 2001, Watts 2001). Intrinsic value refers to a user's value that is independent of the link structure of an online social network. One example of a user's intrinsic value is his or her membership fee paid to LinkedIn (LinkedIn 10-K 2013)[2]. Network value, on the other hand, denotes a user's value that is dependent on the link structure of an online social network. Let us consider advertisements, the major revenue source for operators of online social networks (Facebook 10-K 2013, LinkedIn 10-K 2013), as an example. In this example, a user's network value is the advertisement revenue contributed by the user, which depends on the number of other users that advertisements initiated by the user (e.g., advertisements initially placed on the user's Facebook page) can reach via the link structure of an online social network (Facebook 10-K 2013, LinkedIn 10-K 2013).

---

[2] LinkedIn offers two types of user accounts: basic and premium. While a basic account is free, LinkedIn Inc. charges a membership fee for premium accounts.

In general, a user's network value depends on the size of the user's direct and indirect neighborhood in an online social network (Ballester et al. 2006, Jackson 2008). Intuitively, a user's network value increases as the user has more direct and indirect neighbors (Ballester et al. 2006, Jackson 2008). Moreover, a user's impact on his or her neighbor decays as the distance between them increases (Granovetter 1973, Newman et al. 2002, Jackson 2008). Therefore, a user's network value can be defined as

$$v_j^N = m_j \cdot \sum_{x=1}^{X} \alpha^x |N_{j,x}|. \quad (2)$$

In Equation (2), $N_{j,x}$ represents the set of $x^{th}$ degree neighbors of $w_j$ and $|\cdot|$ denotes the cardinality of a set. For example, $N_{j,1}$ refers to first degree neighbors of $w_j$ or direct neighbors of $w_j$. To model a user's diminishing impact on his or her farther neighbors, we introduce decay factor $\alpha \in (0, 1)$ in Equation (2). Locality parameter $X$ specifies the farthest neighbors considered when calculating a user's network value. The value of $X$ is set such that $\alpha^X |N_{j,X}|$ becomes trivial (Jackson and Rogers 2005, Jackson 2008). Having defined the network impact of $w_j$ as $\sum_{x=1}^{X} \alpha^x |N_{j,x}|$, we use $m_j$ to model the value contribution by one unit of this impact. For example, $m_j$ can be estimated as the revenue generated if an advertisement initiated by $w_j$ reaches a (direct or indirect) neighbor of $w_j$, multiplied by the number of advertisements initiated by $w_j$.

Combining Equations (1) and (2), we can compute a user's value $v_j$ as

$$v_j = v_j^I + m_j \cdot \sum_{x=1}^{X} \alpha^x |N_{j,x}|. \quad (3)$$

For an online social network with $n$ users, the total value $TV$ of these users can be obtained by summing the value of each user in the network (Jackson 2008). We therefore have

$$TV = \sum_{j=1}^{n} v_j. \quad (4)$$

We are now ready to define the value of a potential link. Let $E$ be the set of links currently existing in an online social network. We denote $e_{jh}$ as a potential link that would connect currently unlinked users $w_j$ and $w_h$. Value $V_{jh}$ of potential link $e_{jh}$ can be calculated as

$$V_{jh} = TV_{E\cup\{e_{jh}\}} - TV_E, \tag{5}$$

where $TV_{E\cup\{e_{jh}\}}$ and $TV_E$ denote the total user value in the online social network with and without link $e_{jh}$ respectively, which can be obtained using Equations (4) and (3). By applying Equation (5) to calculate $V_{jh}$, users' intrinsic values before and after adding $e_{jh}$ cancel each other out. This is reasonable because adding a link to an online social network only affects the structure of the network and intrinsic value is independent of network structure.

By recommending potential link $e_{jh}$, value $V_{jh}$ is realized if the recommended link is accepted by users $w_j$ and $w_h$ and hence established. On the other hand, if the recommended link is considered irrelevant by $w_j$ or $w_h$ and thus not established, cost $C_{jh}$ is incurred. One example of $C_{jh}$ is the opportunity cost of not being able to recommend another potential link because of recommending $e_{jh}$, considering that the number of links recommended to a user is limited. Therefore, utility $U_{jh}$ of recommending potential link $e_{jh}$ can be computed as

$$U_{jh} = I_{jh} \cdot V_{jh} - \left(1 - I_{jh}\right) \cdot C_{jh} \ , \tag{6}$$

where $I_{jh} = 1$ if $e_{jh}$ is established and $I_{jh} = 0$ otherwise.

Having defined the utility of recommending a potential link, we can formulate the utility-based link recommendation problem as follows:

*Given an online social network, its users $W$, its existing links $E$,and K, recommend top-K potential links* [3] *with the highest utilities among all potential links, where the utility of recommending a potential link is defined in Equation (6).*

## 4. Utility-based Link Recommendation Method

To solve the utility-based link recommendation problem, we first identify key factors determining utility-based link recommendation decision and construct a Bayesian network to capture dependency relationships among the identified factors and utility-based link recommendation decision. We then propose how to learn the distribution of each identified factor in the Bayesian network and how to predict the probability of recommending a potential link with the learned Bayesian network. Finally, potential links with the highest recommendation probabilities are recommended.

[3] The *K* recommended links are for all users. A user receives a recommendation, if a recommended link has the user as an end point.

**4.1 A Bayesian Network for Utility-based Link Recommendation**

Utility-based link recommendation decision depends on three factors: value ($V$), cost ($C$), and latent linkage likelihood ($L$). The value factor ($V$) refers to the value of a potential link, which can be calculated using Equation (5). The cost factor ($C$) stands for the cost incurred if a potential link is recommended but not established. The latent linkage likelihood factor ($L$) represents the likelihood that a potential link will be established. Link recommendation brings value to an operator, if a recommended link is established, or incurs cost otherwise. Thus, linkage likelihood is an essential factor for utility-based link recommendation decision. Further, linkage likelihood is unobserved; thus it is latent in our proposed Bayesian network. The utility of a potential link depends on its value ($V$), cost ($C$), and linkage likelihood ($L$) and the objective of utility-based link recommendation is to recommend links with the highest utilities. Therefore, factors value ($V$), cost ($C$), and latent linkage likelihood ($L$) jointly determine utility-based link recommendation decision $R$, where $R$ = 1 if a potential link is recommended and $R$ = 0 otherwise.

It has been shown theoretically and empirically that nodal proximity and structural proximity are two effective predictors for linkage likelihood ($L$) (Heider 1958, McPherson et al. 2001, Liben-Nowell and Kleinberg 2007, Crandall et al. 2010). Nodal proximity denotes the similarity between users in terms of their individual characteristics such as age, gender, and education (Chen et al. 2009, Crandall et al. 2010, Schifanella et al. 2010). For two users $w_j$ and $w_h$, nodal proximity $N(w_j, w_h)$ between them is calculated as

$$N(w_j, w_h) = sim(\boldsymbol{r}_j, \boldsymbol{r}_h), \tag{7}$$

where $sim()$ is a similarity function and $\boldsymbol{r}_j$ and $\boldsymbol{r}_h$ denote characteristics of $w_j$ and $w_h$ respectively. Choice of the similarity function depends on data types of user characteristics (Tan et al. 2005); the specific similarity function used in this study is described in §5.2. The effectiveness of nodal proximity in predicting linkage likelihood can be explained by homophily theory, which states that "similarity breeds connection" (McPherson et al. 2001). Therefore, the higher the nodal proximity between users, the more likely a link connecting them will be established.

Structural proximity measures the proximity between users using their structural features in a social network (Liben-Nowell and Kleinberg 2007). Prior studies have empirically shown the power of structural proximity metrics in predicting linkage likelihood (e.g., Liben-Nowell and Kleinberg 2007). Among structural proximity metrics, the Katz index consistently performs well in predicting linkage likelihood (Liben-Nowell and Kleinberg 2007). We thus adopt the Katz index to measure structural proximity between users. Accordingly, the structural proximity $S(w_j, w_h)$ between users $w_j$ and $w_h$ is given by

$$S(w_j, w_h) = \sum_{k} \beta^k \left|path_{jh}^{<k>}\right|, \tag{8}$$

where $path_{jh}^{<k>}$ represents the set of length-$k$ paths connecting users $w_j$ and $w_h$, $|\cdot|$ is the cardinality of a set, and weight $\beta$ is between 0 and 1 (Katz 1953, Liben-Nowell and Kleinberg 2007). The predictive power of the Katz index is rooted in cognitive balance theory (Heider 1958). According to this theory, sentiments (or attitudes) of indirectly connected users could become consistent gradually, which in turn could drive them to link to each other (Heider 1958). In this light, the more paths connecting two users and the shorter the distances of these paths, the more likely it is that a link connecting them will be established.

Having identified key factors underlying utility-based link recommendation decision, we model dependency relationships among them using a Bayesian network. Our choice of Bayesian network is driven by the following considerations. First, Bayesian network is a natural choice for probability prediction and thus we choose it to predict the probability of recommending a potential link. Second, Bayesian network is a powerful but easy to understand model to capture dependencies among variables (Heckerman 2008, Zheng and Pavlou 2010). As shown in Figure 3, the proposed Bayesian network consists of five factors: value ($V$), cost ($C$), structural proximity ($S$), nodal proximity ($N$), and latent linkage likelihood ($L$), as well as utility-based link recommendation decision ($R$). The network assumes mutual independences among factors value (*V*), cost (*C*), structural proximity (*S*), and nodal proximity (*N*); and it captures two dependency relationships: (i) value (*V*), cost (*C*), and latent linkage likelihood ($L$) jointly determine utility-based link recommendation decision ($R$); and (ii) structural proximity (*S*) and nodal proximity (*N*) together predict latent linkage likelihood ($L$).

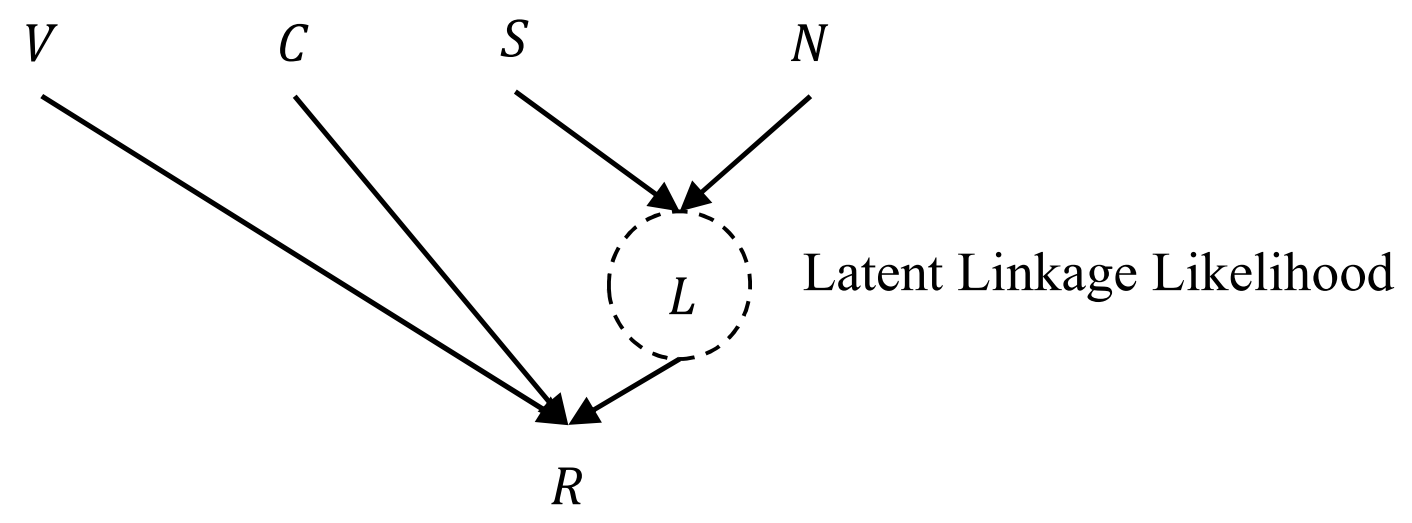

**Figure 3:** A Bayesian Network for Utility-Based Link Recommendation

### 4.2 Learning the Bayesian Network

To employ the Bayesian network to predict recommendation probability, we need to learn the distributions of its factors $V$, $C$, $S$, $N$, and $L$. As a first step of this learning task, we construct training data from observed link establishments in an online social network. Let *t* be current time. To construct training data, we focus on potential links at time $t-1$, i.e., those that had not been established by time $t-1$. For a potential link $i$ at time $t-1$, we can calculate its $V_i$, $N_i$, and $S_i$ according to Equations (5), (7), and (8) respectively and estimate its $C_i$, based on the status[4] of the social network at time $t-1$. It is noted that we can observe whether link $i$ is established or not at current time *t*. Therefore, we set $R_i = 1$ if link $i$ is ranked top-*K* among all potential links in terms of the utility of recommending a potential link defined in Equation (6), and set $R_i = 0$ otherwise. We now have one training record $O_i =< V_i, C_i, S_i, N_i, R_i >$. Continuing the procedure for other potential links, we can construct training data $O = \{O_i\}$, where $i = 1,2,\ldots, M$ and $M$ is the number of records in the training data. We note that linkage likelihood $L$ is unobserved. Thus, we don't have training data on $L$ but we need to learn the distribution of $L$, which is the key methodological challenge for learning the

[4] The status of a social network such as its structure evolves over time.

Bayesian network.

To learn the distributions of factors *V*, *C*, *S*, *N*, and $L$ from training data *O*, we assume exponential family distributions for these factors by following a common Bayesian network learning procedure (Heckerman 2008). Thus, we need to learn parameters of these assumed distributions and denote the vector of these parameters as $\boldsymbol{\theta}$. Specifics about the assumed distributions and their parameters will be discussed later in this subsection. Parameters in $\boldsymbol{\theta}$ can be estimated as those that maximize the log-likelihood $H(O|\boldsymbol{\theta})$ of training data *O* given $\boldsymbol{\theta}$ (Friedman 1998, Mitchell 1997). Formally, the optimal estimation $\boldsymbol{\theta}^*$ of $\boldsymbol{\theta}$ is given by

$$\boldsymbol{\theta}^* = \underset{\boldsymbol{\theta}}{\operatorname{argmax}}\, H(O|\boldsymbol{\theta}), \tag{9}$$

where

$$H(O|\boldsymbol{\theta}) = \sum_{i=1}^{M} \ln[P(O_i|\boldsymbol{\theta})] \tag{10}$$

and $\ln[P(O_i|\boldsymbol{\theta})]$ is the log-likelihood of training record $O_i$ given $\boldsymbol{\theta}$.

However, we cannot obtain $\boldsymbol{\theta}^*$ using Equation (9), because we don't have training data on linkage likelihood *L* but we need to estimate parameters for *L*. To address this challenge, we propose a Bayesian network learning algorithm based on the framework of expectation-maximization (EM), a framework for learning from incomplete data (Dempster et al. 1977). Following the EM framework (Dempster et al. 1977), our algorithm estimates parameters in $\boldsymbol{\theta}$ through an iterative process. In each iteration, our algorithm takes previous parameter estimates as input and produces updated parameter estimates by maximizing the objective function $Q(\cdot)$. Concretely, we have

$$\boldsymbol{\theta}_{k+1} = \underset{\boldsymbol{\theta}}{\operatorname{argmax}}\, Q(\boldsymbol{\theta}|\boldsymbol{\theta}_k), \tag{11}$$

where $\boldsymbol{\theta}_k$ and $\boldsymbol{\theta}_{k+1}$ denote the vector of parameter estimates in iterations $k$ and $k+1$ respectively and $k = 0,1,2,\ldots$. The objective function $Q(\cdot)$ is defined as

$$Q(\boldsymbol{\theta}|\boldsymbol{\theta}_k) = \sum_{i=1}^{M} \int \ln[P(O_i, L_i|\boldsymbol{\theta})] P(L_i|O_i, \boldsymbol{\theta}_k)\, d_{L_i}, \tag{12}$$

where $\ln[P(O_i, L_i|\boldsymbol{\theta})]$ is the log-likelihood of complete data (including observed training data $O_i$ and unobserved linkage likelihood $L_i$) given $\boldsymbol{\theta}$ and $P(L_i|O_i, \boldsymbol{\theta}_k)$ is the probability of $L_i$ given $O_i$ and previous parameter estimates $\boldsymbol{\theta}_k$. In Equation (12), $\int \ln[P(O_i, L_i|\boldsymbol{\theta})]P(L_i|O_i, \boldsymbol{\theta}_k)\, d_{L_i}$ represents the expected log-likelihood of complete data, expected on $L_i$. The iterative process stops if the absolute difference between $H(O|\boldsymbol{\theta}_{k+1})$ and $H(O|\boldsymbol{\theta}_k)$ is sufficiently small. According to Bishop (2006), the iterative process is guaranteed to converge and the converged parameter estimation by our algorithm is a local optimal estimation of $\boldsymbol{\theta}$ for function $H(\cdot)$ defined in Equation (10). For parameter estimation from incomplete data, local optimum is the best result possible (Bishop 2006).

While the above-discussed iterative process of learning $\boldsymbol{\theta}$ follows the standard EM framework (Dempster et al. 1977), the core of the process, i.e., how to compute $\boldsymbol{\theta}_{k+1}$ from $\boldsymbol{\theta}_k$ according to Equations (11) and (12), is specific to our study and represents the key methodological contribution of our proposed algorithm for learning the Bayesian network. In the following, we identify the parameters in $\boldsymbol{\theta}$ by properly decomposing the objective function $Q(\cdot)$ and then show how to compute $\boldsymbol{\theta}_{k+1}$ from $\boldsymbol{\theta}_k$. To decide the specific parameters in $\boldsymbol{\theta}$, we rewrite the objective function $Q(\cdot)$ as

$$\begin{aligned} Q(\boldsymbol{\theta}|\boldsymbol{\theta}_k) = \; & \sum_{i=1}^{M} \ln[P(R_i|\boldsymbol{\theta})] + \sum_{i=1}^{M} \ln[P(V_i|R_i, \boldsymbol{\theta})] + \sum_{i=1}^{M} \ln[P(C_i|R_i, \boldsymbol{\theta})] + \\ & \sum_{i=1}^{M} \int \{\ln[P(S_i, L_i|R_i, \boldsymbol{\theta})] + \ln[P(N_i, L_i|R_i, \boldsymbol{\theta})] - \ln[P(L_i|R_i, \boldsymbol{\theta})]\} P(L_i|O_i, \boldsymbol{\theta}_k) d_{L_i} \end{aligned} \tag{13}$$

The derivation of Equation (13) is given in Appendix A. According to Equation (13), we need to estimate $P(R_i), P(V_i|R_i), P(C_i|R_i), P(S_i, L_i|R_i), P(N_i, L_i|R_i)$, and $P(L_i|R_i)$ for $R_i = 0{,}1$. To estimate $P(R_i)$, we denote parameters $p_0 = P(R_i = 0)$ and $p_1 = P(R_i = 1)$. It is common to assume an exponential family distribution (e.g., exponential or normal) for a continuous factor in a Bayesian network (Friedman 1998, Heckerman 2008). Since factor *V* is continuous and positive, we assume an exponential distribution for factor *V* given *R*. Accordingly, we can estimate $P(V_i|R_i = 0)$ with an exponential density $\lambda_V^0 e^{-\lambda_V^0 \times V_i}$and $P(V_i|R_i = 1)$ with an exponential density $\lambda_V^1 e^{-\lambda_V^1 \times V_i}$. Hence, to estimate $P(V_i|R_i)$, we need to estimate parameter $\lambda_V^{R_i}$ for $R_i = 0{,}1$. In a similar way, we assume factor *C* given *R* following an exponential distribution and estimate $P(C_i|R_i)$ with its density. Thus, to estimate $P(C_i|R_i)$, we need to estimate parameter $\lambda_C^{R_i}$ for $R_i = 0{,}1$.

Similarly, we assume that the joint distribution of factors *S* and *L* given *R* follows a bivariate exponential distribution and estimate $P(S_i, L_i|R_i)$ using its density. In particular, Freund (1961) defines the density function of a bivariate exponential distribution as

$$f(x,y)=\begin{cases}\lambda_x\lambda'_y e^{-\lambda'_y\cdot y-(\lambda_x+\lambda_y-\lambda'_y)\cdot x} & \text{for } 0<x<y,\\ \lambda_y\lambda'_x e^{-\lambda'_x\cdot x-(\lambda_x+\lambda_y-\lambda'_x)\cdot y} & \text{for } 0<y<x.\end{cases} \tag{14}$$

Using the bivariate exponential density defined in Equation (14) to estimate $P(S_i, L_i|R_i)$, we have

$$P(S_i,L_i|R_i)=\begin{cases}\lambda_S^{R_i}\lambda_L'^{R_i}e^{-\lambda_L'^{R_i}\cdot L_i-\left(\lambda_S^{R_i}+\lambda_L^{R_i}-\lambda_L'^{R_i}\right)\cdot S_i} & \text{for } 0<S_i<L_i,\\ \lambda_L^{R_i}\lambda_S'^{R_i}e^{-\lambda_S'^{R_i}\cdot S_i-\left(\lambda_S^{R_i}+\lambda_L^{R_i}-\lambda_S'^{R_i}\right)\cdot L_i} & \text{for } 0<L_i<S_i.\end{cases} \tag{15}$$

We thus need to estimate parameters $\lambda_S^{R_i}$, $\lambda_S'^{R_i}$, $\lambda_L^{R_i}$, and $\lambda_L'^{R_i}$ for $R_i=0,1$. In a similar way, we assume the joint distribution of factors *N* and *L* given *R* following a bivariate exponential distribution and estimate $P(N_i, L_i|R_i)$ as

$$P(N_i,L_i|R_i)=\begin{cases}\lambda_N^{R_i}\lambda_L'^{R_i}e^{-\lambda_L'^{R_i}\cdot L_i-\left(\lambda_N^{R_i}+\lambda_L^{R_i}-\lambda_L'^{R_i}\right)\cdot N_i} & \text{for } 0<N_i<L_i,\\ \lambda_L^{R_i}\lambda_N'^{R_i}e^{-\lambda_N'^{R_i}\cdot N_i-\left(\lambda_N^{R_i}+\lambda_L^{R_i}-\lambda_N'^{R_i}\right)\cdot L_i} & \text{for } 0<L_i<N_i.\end{cases} \tag{16}$$

Hence, we need to estimate parameters $\lambda_N^{R_i}$, $\lambda_N'^{R_i}$, $\lambda_L^{R_i}$, and $\lambda_L'^{R_i}$ for $R_i=0,1$. Finally, we need to estimate $P(L_i|R_i)$. Factor *L* participates in the joint distribution with factor *S* and the joint distribution with factor *N*, whose densities are defined in Equations (15) and (16) respectively. Given *R*, factor *L* follows an exponential distribution with parameter $\lambda_L^R$ if $L<\min(S,N)$ and with parameter $\lambda_L'^R$ otherwise (Freund 1961), where $\min(x,y)$ returns the minimum between *x* and *y*. We thus have

$$P(L_i|R_i)=\begin{cases}\lambda_L^{R_i}e^{-\lambda_L^{R_i}\cdot L_i} & \text{for } L_i<\min(S_i,N_i),\\ \lambda_L'^{R_i}e^{-\lambda_L'^{R_i}\cdot L_i} & \text{for } L_i\geq\min(S_i,N_i).\end{cases} \tag{17}$$

In sum, the parameter vector $\boldsymbol{\theta}$ to be estimated is $\boldsymbol{\theta}=<p_0,\ p_1,\ \lambda_V^0,\ \lambda_V^1,\ \lambda_C^0,\ \lambda_C^1,\ \lambda_S^0,\ \lambda_S^1,\ \lambda_S'^0,\ \lambda_S'^1,\ \lambda_N^0,\ \lambda_N^1,\ \lambda_N'^0,\ \lambda_N'^1,\ \lambda_L^0,\ \lambda_L^1,\ \lambda_L'^0,\ \lambda_L'^1>$. We next show how to compute $\boldsymbol{\theta}$ that maximizes the objective function $Q(\cdot)$ defined in Equation (12).

**Theorem 1.** Given previous parameter estimation $\boldsymbol{\theta}_k=<\bar{p}_0,\ \bar{p}_1,\ \bar{\lambda}_V^0,\ \bar{\lambda}_V^1,\ \bar{\lambda}_C^0,\ \bar{\lambda}_C^1,\ \bar{\lambda}_S^0,\ \bar{\lambda}_S^1,\ \bar{\lambda}_S'^0,\ \bar{\lambda}_S'^1,\ \bar{\lambda}_N^0,\ \bar{\lambda}_N^1,\ \bar{\lambda}_N'^0,\ \bar{\lambda}_N'^1,\ \bar{\lambda}_L^0,\ \bar{\lambda}_L^1,\ \bar{\lambda}_L'^0,\ \bar{\lambda}_L'^1>$ and the exponential distribution assumption for factors *V*, *C*, *S*, *N*, and *L*, there exists a single optimal solution of $\boldsymbol{\theta}$ that maximizes the objective function defined in Equation (12) and the optimal solution is of closed-form.

Proof: See Appendix B for the proof and closed-form solution of $\boldsymbol{\theta}$.

The existence of a single closed-form solution of $\boldsymbol{\theta}$, as discovered by Theorem 1, is an attractive property because a closed-form solution of parameter estimates not only greatly simplifies the implementation of an EM-based algorithm but also considerably improves its computation efficiency (McLachlan and Krishnan 2007). Armed with Theorem 1, we propose an algorithm to learn $\boldsymbol{\theta}$, namely the Bayesian Network Learning with Continuous Latent Factor (BNLF) algorithm. As shown in Figure 4, the algorithm starts with an initial estimation $\boldsymbol{\theta}_0$ of $\boldsymbol{\theta}$ and iteratively updates its estimation according to Theorem 1 until convergence. To obtain $\boldsymbol{\theta}_0$, we follow a common method of parameter initialization, namely repeated random initialization (Duda and Hart 1973). Specifically, we randomly sample one third of the training data and estimate parameters in $\boldsymbol{\theta}_0$ according to the sample. Details of $\boldsymbol{\theta}_0$ estimation are given in Appendix C. We then run the BNLF algorithm with $\boldsymbol{\theta}_0$ and obtain one $\widehat{\boldsymbol{\theta}}$. The above process is repeated for three times. We finally choose $\widehat{\boldsymbol{\theta}}$ that has the largest log-likelihood $H\left(O\middle|\widehat{\boldsymbol{\theta}}\right)$ among the three obtained $\widehat{\boldsymbol{\theta}}$s.

**BNLF** $(O, \epsilon)$
    $O$: training data
    $\epsilon$: predefined convergence threshold

Initialize $\boldsymbol{\theta}_0$. //$\boldsymbol{\theta}_0$: initial estimation of $\boldsymbol{\theta}$
$k = -1$.
**Do**
    $k = k + 1$.
    Obtain $\boldsymbol{\theta}_{k+1}$ according to Equation (11) and Theorem 1.
**While** $(|H(O|\boldsymbol{\theta}_{k+1}) - H(O|\boldsymbol{\theta}_k)| > \epsilon)$ // $H(\cdot)$: log-likelihood defined in Equation (10)
$\widehat{\boldsymbol{\theta}} = \boldsymbol{\theta}_{k+1}$. //$\widehat{\boldsymbol{\theta}}$: final estimation of $\boldsymbol{\theta}$
**Return** $\widehat{\boldsymbol{\theta}}$.

**Figure 4:** The BNLF Algorithm for Learning $\boldsymbol{\theta}$

### 4.3 Predicting Recommendation Probability

Having obtained parameter estimation $\widehat{\boldsymbol{\theta}} =< \hat{p}_0,\ \hat{p}_1,\ \hat{\lambda}_V^0,\ \hat{\lambda}_V^1,\ \hat{\lambda}_C^0,\ \hat{\lambda}_C^1,\ \hat{\lambda}_S^0,\ \hat{\lambda}_S^1,\ \hat{\lambda}_S'^0,\ \hat{\lambda}_S'^1,\ \hat{\lambda}_N^0,\ \hat{\lambda}_N^1,\ \hat{\lambda}_N'^0,\ \hat{\lambda}_N'^1,\ \hat{\lambda}_L^0,\ \hat{\lambda}_L^1,\ \hat{\lambda}_L'^0,\ \hat{\lambda}_L'^1 >$, we are ready to predict the recommendation probability for each potential link. In an online social network, for a potential link $a$ at current time $t$, i.e., a link that has not been established by current time $t$, we can calculate its $V_a$, $N_a$, and $S_a$ according to Equations (5), (7), and (8) respectively and estimate its $C_a$, based on the status of the network at current time $t$. We define the recommendation probability for potential link $a$ as the probability of recommending it given its $V_a$, $C_a$, $S_a$, $N_a$, and parameter estimation $\hat{\boldsymbol{\theta}}$, i.e., $P(R_a = 1|V_a, C_a, S_a, N_a, \hat{\boldsymbol{\theta}})$. In Appendix D, we show how to compute recommendation probability $P(R_a = 1|V_a, C_a, S_a, N_a, \hat{\boldsymbol{\theta}})$**.** Applying the formula for computing recommendation probability given in Appendix D, we can predict recommendation probability for each potential link and recommend top-$K$ potential links with the highest recommendation probabilities.

## 5. Empirical Evaluation

We conducted experiments to evaluate our method using real-world social network data. In this section, we describe the data and parameter calibration, detail our experimental procedure, evaluation metrics, and benchmark methods, and report experimental results.

### 5.1 Data and Parameter Calibration

We collected data from a major U.S. online social network over a one-year period, starting from the launch of the network. One collected data set describes who registered on which date as a user of the online social network; another data set contains data on who is linked to whom and when the linkage was established. As shown in Figure 5, both the number of users and the number of links grow rapidly over time. At the end of the one-year period, the online social network had 485,608 users, connected by 669,524 links. For each user, we also collected data about his or her profile. Because of privacy concerns, the profile of a user consists of a set of encoded terms, each of which corresponds to a characteristic of the user. During the one-year period, no link recommendation mechanism had been deployed in the online social network. Thus, our data provide a natural test bed for evaluating different link recommendation methods.

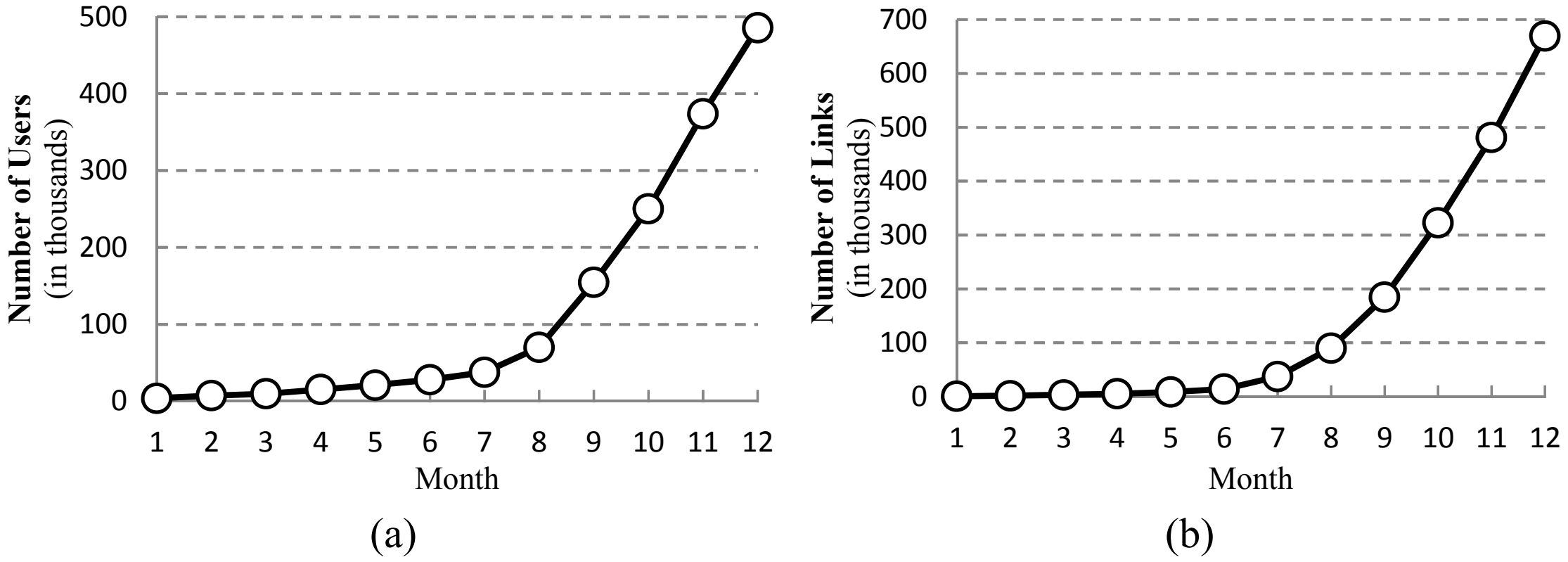

**Figure 5:** The Growth of Users (a) and Links (b) in the Online Social Network

We then calibrated parameters for the utility-based link recommendation problem. Following Jackson (2008), we set decay factor $\alpha$ in Equation (2) to 0.5. Parameter $m_j$ in Equation (2) was estimated as the revenue generated if an advertisement initiated by user $w_j$ reaches a (direct or indirect) neighbor of $w_j$, multiplied by the average number of advertisements initiated by a user in a month.[5] Cost $C_{jh}$ in Equation (6) was initially treated as the opportunity cost of not being able to recommend another potential link to user $w_j$ or $w_h$ because of recommending $e_{jh}$. We thus estimated $C_{jh}$ as the average value of links actually established by user $w_j$ or $w_h$, where the value of a link was computed using Equation (5). For robustness analysis, we conducted additional experiments with $\rho \times$ intial cost estimation, where $\rho = 0.5, 2$. Following a common practice in link recommendation (Backstrom and Leskovec 2011, Wang et al. 2011, Dong et al. 2012), we focus on potential links that, if established, would connect users who are two hops away. Compared to considering all possible potential links (i.e., potential links that would connect users two or more hops away), focusing on potential links that would connect users two hops away can greatly accelerate computation without sacrificing much on prediction (Backstrom and Leskovec 2011, Wang et al. 2011, Dong et al. 2012). In our experiments, the number of potential links that would connect users two hops away increases over time because the number of users increases over time[6]. Hence, rather than setting $K$ to a static number, we set $K$ as a percentage of potential links in a month. We therefore set $K$=0.5%× number of potential links in a month, i.e., recommending top 0.5% of potential links in a month. To ensure the robustness of our empirical evaluation, we further conducted experiments with $K$=0.25%× number of potential links in a month and with $K$=0.75%× number of potential links in a month.

---

[5] Due to privacy concerns, we do not have data on the number of advertisements initiated by a specific user but only the average number across users.

[6] The number of potential links that would connect users two hops away is as follows: 110475 (1), 176216 (2), 252963 (3), 483948 (4), 709991 (5), 931623 (6), 1333526 (7), 3291007 (8), 8808065 (9), 15988577 (10), 26317250 (11), and 38517866 (12). Here, month number is enclosed in the parentheses.

### 5.2 Experimental Design

Our experiments follow the procedure described below. Let month $t$ be the current month and month $t+1$ be the prediction month. We use data by current month $t$ to train our method as well as each benchmark method to recommend top-$K$ potential links[7] out of potential links in month $t$. Using data by prediction month $t+1$, we can verify whether a potential link in month $t$ is actually accepted and established by users in month $t+1$, compute its utility with Equation (6), and identify true top-$K$ potential links that have the highest utilities. The performance of a method is then evaluated using *top-K utility-based precision*, which is the fraction of recommended top-$K$ potential links that are true top-$K$ potential links. In addition, we also evaluate the *average utility* of links recommended by a method. After all, the objective of the utility-based link recommendation problem is to recommend links with the highest utilities. Therefore, a method that better achieves the objective recommends links with higher average utility.

We next discuss implementation details of our method and benchmark methods. For our method, nodal proximity between users was measured using Equation (7), for which we chose the Jaccard coefficient as the similarity function. In our experiments, user profile is represented by a set of terms and the Jaccard coefficient is suitable for measuring similarity between sets (Salton and McGill 1983). Specifically, the Jaccard coefficient is defined as

$$sim(\boldsymbol{r}_j, \boldsymbol{r}_h) = \frac{|\boldsymbol{r}_j \cap \boldsymbol{r}_h|}{|\boldsymbol{r}_j \cup \boldsymbol{r}_h|} \tag{19}$$

where $\boldsymbol{r}_j$ and $\boldsymbol{r}_h$ denote the set of profile terms of users $w_j$ and $w_h$ respectively and $|\cdot|$ is the cardinality of a set. Structural proximity between users was computed using Equation (8). Following a common practice (Liben-Nowell and Kleinberg 2007, Lichtenwalter et al. 2010), we set $\beta$ in Equation (8) to 0.05.

[7] The $K$ recommended links are for all users. By simply adjusting its output, our method can target a user and recommend the same number of links (say $m$ links) to each user. Specifically, for a user, our method can be adapted to identify and recommend $m$ potential links that have the highest recommendation probabilities among potential links with this user as an end point.

We selected a representative method from each category of existing link recommendation methods as a benchmark. For the category of learning-based methods, we chose the SVM-based link recommendation method because of its outstanding performance among learning-based methods (Hasan et al. 2006). Our implementation of the SVM-based method followed its implementation described in (Hasan et al. 2006, Lichtenwalter et al. 2010). Among structural proximity-based methods, the Katz index was selected due to its superior performance (Liben-Nowell and Kleinberg 2007). A suitable nodal proximity-based method for our experiments is the Jaccard coefficient (i.e., Equation (19)), thus chosen as a benchmark. Moreover, we benchmarked our method against a link recommendation method commonly used in practice – common neighbor[8]. In particular, the common neighbor $CN_{jh}$ between users $w_j$ and $w_h$ is computed as the number of their mutual neighbors (Newman 2001, Liben-Nowell and Kleinberg 2007):

$$CN_{jh} = |\Gamma_j \cap \Gamma_h| \tag{20}$$

where $\Gamma_j$ and $\Gamma_h$ denote the set of direct neighbors of users $w_j$ and $w_h$ respectively. We also benchmarked our method against a variation of the common neighbor – Adamic/Adar (Adamic and Adar 2003). Extended from the common neighbor, the Adamic/Adar $AA_{jh}$ between users $w_j$ and $w_h$ is given by (Adamic and Adar 2003, Liben-Nowell and Kleinberg 2007)

$$AA_{jh} = \sum_{w_z \in \Gamma_j \cap \Gamma_h} \frac{1}{\log |\Gamma_z|} \tag{21}$$

where $\Gamma_j$ , $\Gamma_h$, and $\Gamma_z$ denote the set of direct neighbors of users $w_j$, $w_h$, and $w_z$ respectively, and $w_z$ is a common neighbor of users $w_j$ and $w_h$. Table 1 summarizes the benchmark methods.

[8] Common neighbor is popularly used by major online social networks for link recommendation. For example, it is called “mutual friend” in Facebook and “shared connection” in LinkedIn.

| Method | Category | Abbreviation |
|---|---|---|
| SVM-based link recommendation | Learning | SVM |
| Katz index | Structural proximity | Katz |
| Jaccard coefficient | Nodal proximity | Jaccard |
| Common neighbor | Structural proximity | CN |
| Adamic/Adar | Structural proximity | AA |

**Table 1:** Summary of Benchmark Methods

### 5.3 Experimental Results and Analysis

Following the experimental procedure, we conducted experiments with the parameter values set in §5.1 and current month $t$ = 2,3,…, 11.[9] As shown in Table 2, our method substantially outperforms each benchmark method in every prediction month, in terms of top-$K$ utility-based precision. Averaged across prediction months, the mean top-$K$ utility-based precision of our method is 0.40, which indicates that on average 40% of top-$K$ potential links recommended by our method are true top-$K$ links. On the other hand, the mean top-$K$ utility-based precision of SVM, the best performing benchmark method in terms of top-$K$ utility-based precision, is only 0.27. On average, the top-$K$ utility-based precision of our method is 45.52% higher than that of SVM and 122.25% higher than that of CN, a link recommendation method commonly used in practice. The outperformance of our method over benchmark methods in terms of top-$K$ utility-based precision is due to its consideration of utility factors such as value ($V$) and cost ($C$) as well as its better link prediction accuracy. Here, the link prediction accuracy of a method is the fraction of potential links recommended by this method that are actually established. For example, averaged across prediction months, our method outperforms SVM by 24.11% in terms of link prediction accuracy.

[9] To construct training data for our method, we need data in month $t-1$ and month $t$. Thus, current month $t$ in our experiments starts from month 2 instead of month 1.

| Prediction Month (*t*+1) | AA | CN | Jaccard | Katz | SVM | Our Method |
|---|---|---|---|---|---|---|
| 3 | 0.18 | 0.17 | 0.05 | 0.25 | 0.28 | 0.46 |
| 4 | 0.19 | 0.20 | 0.04 | 0.27 | 0.31 | 0.42 |
| 5 | 0.16 | 0.16 | 0.05 | 0.20 | 0.26 | 0.36 |
| 6 | 0.16 | 0.20 | 0.03 | 0.23 | 0.27 | 0.41 |
| 7 | 0.17 | 0.16 | 0.03 | 0.22 | 0.26 | 0.37 |
| 8 | 0.18 | 0.18 | 0.03 | 0.25 | 0.25 | 0.35 |
| 9 | 0.16 | 0.17 | 0.03 | 0.22 | 0.25 | 0.37 |
| 10 | 0.16 | 0.16 | 0.02 | 0.22 | 0.24 | 0.39 |
| 11 | 0.19 | 0.20 | 0.02 | 0.27 | 0.29 | 0.45 |
| 12 | 0.20 | 0.21 | 0.03 | 0.28 | 0.30 | 0.39 |
| Mean | 0.17 | 0.18 | 0.03 | 0.24 | 0.27 | 0.40 |
| Std. | 0.01 | 0.02 | 0.01 | 0.03 | 0.02 | 0.04 |

**Table 2:** Top-*K* Utility-based Precision: Our Method versus Benchmark Methods

Table 3 compares the average utility of links recommended by our method against that by each benchmark method. As shown, our method significantly outperforms each benchmark method across prediction months, in terms of average utility. Averaged across prediction months, our method outperforms SVM (the best performing benchmark method in terms of average utility) by 41.76% and CN (a method commonly used in practice) by 210.46%. For example, in prediction month 12, the average utility of links recommended by our method is $1.87, which means that on average a link recommended by our method could bring $1.87 to the operator of the online social network. In comparison, the average utility of links recommended by SVM is $1.56 and the average utility of links recommended by CN is $0.71. In other words, a link recommended by our method could generate an average of $0.31 more revenue than a link recommended by SVM and an average of $1.16 more revenue than a link recommended by CN. Considering the sheer number of links recommended in the online social network, successful implementation of our method could bring significant financial gains to the network's operator, compared to benchmark methods.

| Prediction Month ($t$+1) | AA | CN | Jaccard | Katz | SVM | Our Method |
|---|---|---|---|---|---|---|
| 3 | $0.29 | $0.21 | $0.10 | $0.44 | $0.92 | $1.15 |
| 4 | $0.44 | $0.47 | $0.12 | $0.71 | $0.82 | $1.14 |
| 5 | $0.27 | $0.25 | $0.06 | $0.32 | $0.76 | $0.99 |
| 6 | $0.30 | $0.33 | $0.07 | $0.36 | $0.51 | $0.83 |
| 7 | $0.28 | $0.26 | $0.04 | $0.39 | $0.54 | $0.78 |
| 8 | $0.37 | $0.38 | $0.05 | $0.55 | $0.56 | $0.97 |
| 9 | $0.36 | $0.37 | $0.05 | $0.52 | $0.74 | $0.99 |
| 10 | $0.37 | $0.38 | $0.04 | $0.55 | $0.74 | $1.17 |
| 11 | $0.60 | $0.64 | $0.06 | $0.99 | $1.35 | $1.77 |
| 12 | $0.65 | $0.71 | $0.06 | $1.07 | $1.56 | $1.87 |
| Mean | $0.39 | $0.40 | $0.07 | $0.59 | $0.85 | $1.17 |
| Std. | $0.13 | $0.16 | $0.02 | $0.26 | $0.35 | $0.37 |

**Table 3:** Average Utility: Our Method versus Benchmark Methods

The performance improvement by our method is attributed to its methodological design. Our method is designed to recommend a link based on its value, cost, and linkage likelihood, essential factors determining the utility of a link, whereas benchmark methods recommend a link on the basis of its linkage likelihood only. As a result, links recommended by our method generally have higher utilities than those recommended by benchmark methods. While it is necessary to consider a link's value and cost when making link recommendation decision, it is also essential to take into account its linkage likelihood. In our method, linkage likelihood is treated as a latent factor. Without this latent factor, a simple solution to the utility-based link recommendation problem is a Naïve Bayes (NB) method, which predicts link recommendation decision ($R$) from input factors: value ($V$), cost ($C$), structural proximity ($S$), and nodal proximity ($N$). Table 4 compares the performance of our method against that of NB. Averaged across prediction months, our method outperforms NB by 80.38% in terms of top-$K$ utility-based precision and 20.22% in terms of average utility. Such improvements echo the necessity of linkage likelihood in link recommendation decision.

| Prediction Month ($t$+1) | Top-$K$ Utility-based Precision | | Average Utility | |
|---|---|---|---|---|
| | **NB** | **Our Method** | **NB** | **Our Method** |
| 3 | 0.31 | 0.46 | $1.04 | $1.15 |
| 4 | 0.26 | 0.42 | $1.02 | $1.14 |
| 5 | 0.21 | 0.36 | $0.82 | $0.99 |
| 6 | 0.19 | 0.41 | $0.63 | $0.83 |
| 7 | 0.18 | 0.37 | $0.61 | $0.78 |
| 8 | 0.19 | 0.35 | $0.80 | $0.97 |
| 9 | 0.18 | 0.37 | $0.83 | $0.99 |
| 10 | 0.20 | 0.39 | $0.91 | $1.17 |
| 11 | 0.25 | 0.45 | $1.48 | $1.77 |
| 12 | 0.23 | 0.39 | $1.70 | $1.87 |
| Mean | 0.22 | 0.40 | $0.98 | $1.17 |
| Std. | 0.04 | 0.04 | $0.35 | $0.37 |

**Table 4:** Performance Comparison between Our Method and NB

In summary, our empirical results show that our method substantially outperforms representative link recommendation methods from prior research. To ensure the robustness of our empirical findings, we conducted experiments with different cost estimations or *K*; experimental results reported in Appendix E further confirm the performance advantage of our method over benchmark methods. We also show the contribution of each component of our method to its performance in Appendix F and demonstrate the outperformance of our method over benchmark methods using another social network data set in Appendix G.

## 6. Conclusions

Link recommendation is a key functionality offered by major online social networks. Existing methods for link recommendation focus on the likelihood of linkage but overlook the benefit of linkage. Our study addresses this limitation and contributes the innovative idea of utility-based link recommendation to extant literature. First, we define the utility of recommending a potential link and formulate a new link recommendation problem. Second, we propose a novel utility-based link recommendation method that recommends links based on the value, cost, and linkage likelihood of a potential link, in contrast to existing link recommendation methods that focus solely on linkage likelihood. Our empirical evaluation demonstrates the performance advantage of our proposed method over prevalent link recommendation methods found in representative prior research. Third, our study also contributes a novel problem and method to the utility-based data mining literature (Weiss et al. 2008, Saar-Tsechansky et al. 2009).

Our study has several managerial implications. First, our research sheds light on the critical role of balancing operator's benefit and users' needs in link recommendation. After all, a potential link can benefit an operator only after it is established by users. In this vein, the effectiveness of link recommendation depends on the proper consideration of the value, cost, and linkage likelihood of a potential link. Failing to consider any of the three factors greatly reduces the effectiveness of a link recommendation method, as demonstrated in our empirical analysis. Our method innovatively integrates these three factors in link recommendation and offers great value to significant applications. For example, we show that on average, a link recommended by our method can produce 41.76% more utility than a link recommended by the best performing benchmark method. Given that advertisement alone is estimated to generate $12 billion revenue to operators of online social networks in 2014 (eMarketer.com 2012), such improvement could create huge financial gains for them.

One important reason our method yields better performance over benchmark methods is the consideration of the network value that a user brings to an operator. In an online social network, users are neither isolated nor independent. On the contrary, they are connected and influence each other. As a result, a user's network value arises from his or her influence on his or her direct and indirect neighbors in a network. Since the major revenue source for an operator comes from advertisement, and pay-per-impression as well as pay-per-click continue to be prevalent business models for online social networks, network value is critical to the business performance of these networks. Therefore, the operator of an online social network should treat network value as a key factor in crucial decisions such as link recommendation and targeted marketing.

Third, our study highlights the important role of linkage likelihood in determining utilities of recommended links. In this light, the operator of an online social network could purposefully boost linkage likelihood such that both the operator and users are benefited. For example, an operator can enhance features that accentuate nodal and structural similarity between users, two factors that jointly determine linkage likelihood. This in turn enables a user to be more aware of other users who are similar to the user, which generally increases the likelihood of link formation. Another promising approach is to provide incentives to facilitate the establishment of potential links. Rather than passively waiting for users to make connections, an operator can actively intervene and provide incentives to lure users to connect. Such approach could be especially useful for potential links with moderate linkage likelihood but high value.

Our research has limitations that should be addressed in future research endeavors. In our Bayesian network model, we assume mutual independences among factors value ($V$), cost ($C$), structural proximity ($S$), and nodal proximity ($N$). Future work should examine how to relax this assumption. One viable approach is to learn dependency relationships among these factors from data (Heckerman 2008). Such an approach could improve the predictive effectiveness of our method although it might increase its computational complexity. In addition, we assume that a better connected social network is more effective in facilitating information diffusion over the network. Future work should also consider users' susceptibilities to information diffused over a network. Moreover, we assume that effective information diffusion over a social network is beneficial for the network's operator and study the positive side of link recommendation, i.e., the utility of link recommendation. Future work should examine the negative aspect of link recommendation, such as the rapid spread of negative sentiments about a product over a more connected social network due to link recommendation.

Additionally, there are research questions worthy of future exploration. First, it is interesting to study how to recommend a set of potential links that collectively has the highest utility. Second, the effectiveness of the learned Bayesian network in predicting recommendation probabilities could decline over time because it is learned from previous user linkage behaviors and does not capture new user linkage behaviors. Thus, another interesting question is how to maintain the currency of the learned Bayesian network over time. Prior research on knowledge refreshing and maintenance (Bensoussan et al. 2009, Fang et al. 2013b) could provide theoretical foundations for this question. Another area worthy of future investigation is to characterize potential links that can increase the value of a social network the most, where value increase due to the addition of a potential link is defined in Equation (5). Some recent exploration in this direction has found that a potential link that reduces the clustering coefficient of a social network could increase the value of the network (Zhao et al. 2012). Such finding could be used in combination with linkage likelihood predicted by an existing link prediction method to recommend links with high utilities. It is also worthwhile to conduct field experiments to evaluate our method. In an experiment, we can observe in real time how users react to recommended links, which recommended links they actually establish, and the values of these established links. By combining evaluation results with archival data and field experimental results, we could produce more comprehensive evidence on the effectiveness of our method. Finally, it is interesting to extend our method by considering a user's historical link adoption record, which documents all other users to whom the user has already linked, i.e., direct neighbors of the user. Understandably, the more similar user $w_j$ to user $w_h$'s direct neighbors, the more likely $w_j$ is the kind of person that $w_h$ likes to connect and the higher the linkage likelihood between $w_j$ and $w_h$ is. Therefore, future work needs to incorporate the similarities between a user and each of another user's direct neighbors into our method.

**Appendix A: Derivation of Equation (13)**

We repeat Equation (12):

$$Q(\boldsymbol{\theta}|\boldsymbol{\theta}_k) = \sum_{i=1}^{M} \int \ln[P(O_i, L_i|\boldsymbol{\theta})]P(L_i|O_i, \boldsymbol{\theta}_k)\, d_{L_i}. \tag{A1}$$

The joint probability $P(O_i, L_i|\boldsymbol{\theta})$ in Equation (A1) can be expanded as

$$\begin{aligned} P(O_i, L_i|\boldsymbol{\theta}) &= P(V_i, C_i, S_i, N_i, L_i, R_i|\boldsymbol{\theta}) \\ &= P(V_i, C_i, L_i, S_i, N_i|R_i, \boldsymbol{\theta})P(R_i|\boldsymbol{\theta}). \end{aligned} \tag{A2}$$

Prior research focuses on learning Bayesian Networks with discrete latent (hidden) variables (e.g., Friedman 1997). However, linkage likelihood in the focal problem is a continuous latent variable, which is the major methodological obstacle for our study. To overcome this obstacle, we need to effectively decompose (A2); accordingly, we assume mutual independences among $V, C, \{S, N, L\}$ given $R$ and the independence between $S, N$ given $L$. These assumptions do not affect the performance of our proposed method much, as demonstrated by the substantial outperformance of our method over benchmark methods in §5. With the assumptions, we can rewrite Equation (A2) as

$$P(O_i, L_i|\boldsymbol{\theta}) = P(V_i|R_i, \boldsymbol{\theta})P(C_i|R_i, \boldsymbol{\theta})P(L_i, S_i, N_i|R_i, \boldsymbol{\theta})P(R_i|\boldsymbol{\theta}). \tag{A3}$$

The last two terms in Equation (A3) can be rewritten as

$$\begin{aligned} P(L_i, S_i, N_i|R_i, \boldsymbol{\theta})P(R_i|\boldsymbol{\theta}) &= P(S_i, N_i|L_i, R_i, \boldsymbol{\theta})P(L_i|R_i, \boldsymbol{\theta})P(R_i|\boldsymbol{\theta}) \\ &= P(S_i|L_i, R_i, \boldsymbol{\theta})P(N_i|L_i, R_i, \boldsymbol{\theta})P(L_i|R_i, \boldsymbol{\theta})P(R_i|\boldsymbol{\theta}) \\ &= \frac{P(S_i, L_i|R_i, \boldsymbol{\theta})P(N_i, L_i|R_i, \boldsymbol{\theta})P(R_i|\boldsymbol{\theta})}{P(L_i|R_i, \boldsymbol{\theta})}. \end{aligned} \tag{A4}$$

Substituting the last two terms in (A3) with (A4), we have

$$P(O_i, L_i|\boldsymbol{\theta}) = \frac{P(V_i|R_i, \boldsymbol{\theta})P(C_i|R_i, \boldsymbol{\theta})P(S_i, L_i|R_i, \boldsymbol{\theta})P(N_i, L_i|R_i, \boldsymbol{\theta})P(R_i|\boldsymbol{\theta})}{P(L_i|R_i, \boldsymbol{\theta})}. \tag{A5}$$

Substituting $P(O_i, L_i|\boldsymbol{\theta})$ in (A1) with (A5), we can obtain Equation (13).

**Appendix B:** Proof of Theorem 1

**Theorem 1.** Given previous parameter estimation $\boldsymbol{\theta}_k =< \bar{p}_0, \bar{p}_1, \bar{\lambda}_V^0, \bar{\lambda}_V^1, \bar{\lambda}_C^0, \bar{\lambda}_C^1, \bar{\lambda}_S^0, \bar{\lambda}_S^1, \bar{\lambda}_S'^0, \bar{\lambda}_S'^1, \bar{\lambda}_N^0, \bar{\lambda}_N^1, \bar{\lambda}_N'^0, \bar{\lambda}_N'^1, \bar{\lambda}_L^0, \bar{\lambda}_L^1, \bar{\lambda}_L'^0, \bar{\lambda}_L'^1 >$ and the exponential distribution assumption for factors *V*, *C*, *S*, *N*, and *L*, there exists a single optimal solution of $\boldsymbol{\theta}$ that maximizes the objective function defined in Equation (12) and the optimal solution given below is of closed-form:

$$p_1 = \frac{\sum_i R_i}{M}, \tag{B1}$$

$$p_0 = \frac{\sum_i (1 - R_i)}{M}, \tag{B2}$$

$$\lambda_V^1 = \frac{\sum_i R_i}{\sum_i R_i V_i}, \tag{B3}$$

$$\lambda_V^0 = \frac{\sum_i (1 - R_i)}{\sum_i (1 - R_i) V_i}, \tag{B4}$$

$$\lambda_C^1 = \frac{\sum_i R_i}{\sum_i R_i C_i}, \tag{B5}$$

$$\lambda_C^0 = \frac{\sum_i (1 - R_i)}{\sum_i (1 - R_i) C_i}, \tag{B6}$$

$$\lambda_S^1 = \frac{\sum_i R_i\,(2 - I_i)}{\sum_i R_i [S_i + \Gamma_i^2 + I_i \Gamma_i^3 + (1 - I_i) S_i]}, \tag{B7}$$

$$\lambda_S^0 = \frac{\sum_i (1 - R_i)(2 - I_i)}{\sum_i (1 - R_i)[S_i + \Gamma_i^2 + I_i \Gamma_i^3 + (1 - I_i) S_i]}, \tag{B8}$$

$$\lambda_S'^1 = \frac{\sum_i R_i\,(1 + I_i)}{\sum_i R_i (S_i - \Gamma_i^2 + I_i S_i - I_i \Gamma_i^3)}, \tag{B9}$$

$$\lambda_S'^0 = \frac{\sum_i (1 - R_i)(1 + I_i)}{\sum_i (1 - R_i)(S_i - \Gamma_i^2 + I_i S_i - I_i \Gamma_i^3)}, \tag{B10}$$

$$\lambda_N^1 = \frac{\sum_i R_i\,(1 + I_i)}{\sum_i R_i [N_i + \Gamma_i^2 + (1 - I_i)\Gamma_i^4 + I_i N_i]}, \tag{B11}$$

$$\lambda_N^0 = \frac{\sum_i (1 - R_i)\,(1 + I_i)}{\sum_i (1 - R_i)[N_i + \Gamma_i^2 + (1 - I_i)\Gamma_i^4 + I_i N_i]}, \tag{B12}$$

$$\lambda_N'^1 = \frac{\sum_i R_i\,(2 - I_i)}{\sum_i R_i [N_i - \Gamma_i^2 + (1 - I_i) N_i - (1 - I_i)\Gamma_i^4]}, \tag{B13}$$

$$\lambda_N'^0 = \frac{\sum_i (1 - R_i)\,(2 - I_i)}{\sum_i (1 - R_i)[N_i - \Gamma_i^2 + (1 - I_i) N_i - (1 - I_i)\Gamma_i^4]}, \tag{B14}$$

$$\lambda_L^1 = \frac{\sum_i R_i\,[1 + I_i \Gamma_i^3 + (1 - I_i)\Gamma_i^4]}{\sum_i R_i [(N_i + S_i) + \Gamma_i^2 + I_i\left(\Gamma_i^3 + N_i\right) + (1 - I_i)(\Gamma_i^4 + S_i)]}, \tag{B15}$$

$$\lambda_L^0 = \frac{\sum_i (1 - R_i)\,[1 + I_i \Gamma_i^3 + (1 - I_i)\Gamma_i^4]}{\sum_i (1 - R_i)[(N_i + S_i) + \Gamma_i^2 + I_i\left(\Gamma_i^3 + N_i\right) + (1 - I_i)(\Gamma_i^4 + S_i)]}, \tag{B16}$$

$$\lambda_L'^1 = \frac{\sum_i R_i}{\sum_i R_i[\Gamma_i^1 - (S_i + N_i) + I_i N_i + (1 - I_i) S_i]}, \tag{B17}$$

$$\lambda_L'^0 = \frac{\sum_i (1 - R_i)}{\sum_i (1 - R_i)[\Gamma_i^1 - (S_i + N_i) + I_i N_i + (1 - I_i) S_i]}, \tag{B18}$$

where

$$I_i = \begin{cases} 1 & \text{if } S_i > N_i, \\ 0 & \text{otherwise,} \end{cases}$$

$$Y_i = \max(S_i, N_i),$$

$$y_i = \min(S_i, N_i),$$

$$\Gamma_i^1 = Y_i + \frac{1}{\bar{\lambda}_L'^{R_i}},$$

$$\Gamma_i^2 = \frac{1}{\bar{\lambda}_S^{R_i} - \bar{\lambda}_S'^{R_i} + \bar{\lambda}_N^{R_i} - \bar{\lambda}_N'^{R_i} + \bar{\lambda}_L^{R_i}} - \frac{y_i \cdot e^{-\left(\bar{\lambda}_S^{R_i} - \bar{\lambda}_S'^{R_i} + \bar{\lambda}_N^{R_i} - \bar{\lambda}_N'^{R_i} + \bar{\lambda}_L^{R_i}\right) y_i}}{1 - e^{-\left(\bar{\lambda}_S^{R_i} - \bar{\lambda}_S'^{R_i} + \bar{\lambda}_N^{R_i} - \bar{\lambda}_N'^{R_i} + \bar{\lambda}_L^{R_i}\right) y_i}},$$

$$\Gamma_i^3 = \frac{N_i \cdot e^{-\left(\bar{\lambda}_S^{R_i} + \bar{\lambda}_L^{R_i} - \bar{\lambda}_S'^{R_i}\right) N_i} - S_i \cdot e^{-\left(\bar{\lambda}_S^{R_i} + \bar{\lambda}_L^{R_i} - \bar{\lambda}_S'^{R_i}\right) S_i}}{e^{-\left(\bar{\lambda}_S^{R_i} + \bar{\lambda}_L^{R_i} - \bar{\lambda}_S'^{R_i}\right) N_i} - e^{-\left(\bar{\lambda}_S^{R_i} + \bar{\lambda}_L^{R_i} - \bar{\lambda}_S'^{R_i}\right) S_i}} + \frac{1}{\bar{\lambda}_S^{R_i} + \bar{\lambda}_L^{R_i} - \bar{\lambda}_S'^{R_i}} \text{, and}$$

$$\Gamma_i^4 = \frac{S_i \cdot e^{-\left(\bar{\lambda}_N^{R_i} + \bar{\lambda}_L^{R_i} - \bar{\lambda}_N'^{R_i}\right) N_i} - N_i \cdot e^{-\left(\bar{\lambda}_N^{R_i} + \bar{\lambda}_L^{R_i} - \bar{\lambda}_N'^{R_i}\right) S_i}}{e^{-\left(\bar{\lambda}_N^{R_i} + \bar{\lambda}_L^{R_i} - \bar{\lambda}_N'^{R_i}\right) N_i} - e^{-\left(\bar{\lambda}_N^{R_i} + \bar{\lambda}_L^{R_i} - \bar{\lambda}_N'^{R_i}\right) S_i}} + \frac{1}{\bar{\lambda}_N^{R_i} + \bar{\lambda}_L^{R_i} - \bar{\lambda}_N'^{R_i}}.$$

**Note:** In this appendix, summation $\sum_i$ denotes the sum from $i = 1$ to $i = M$.

**Proof.** We first derive a closed-form optimal solution of $\boldsymbol{\theta}$ and then show the derived solution is a single optimal solution of $\boldsymbol{\theta}$ that maximizes the objective function $Q(\cdot)$ defined in Equation (12). According to Equation (13), the objective function $Q(\cdot)$ can be written as

$$\begin{aligned} Q(\boldsymbol{\theta}|\boldsymbol{\theta}_k) = & \sum_{i=1}^{M} \ln[P(R_i|\boldsymbol{\theta})] + \sum_{i=1}^{M} \ln[P(V_i|R_i, \boldsymbol{\theta})] + \sum_{i=1}^{M} \ln[P(C_i|R_i, \boldsymbol{\theta})] + \\ & \sum_{i=1}^{M} \int \{\ln[P(S_i, L_i|R_i, \boldsymbol{\theta})] + \ln[P(N_i, L_i|R_i, \boldsymbol{\theta})] - \ln[P(L_i|R_i, \boldsymbol{\theta})]\} P(L_i|O_i, \boldsymbol{\theta}_k) d_{L_i}. \end{aligned} \tag{B19}$$

The last term in Equation (B19) can be rewritten as

$$\begin{aligned} P(L_i|O_i, \boldsymbol{\theta}_k) = & \; P(L_i|V_i, C_i, S_i, N_i, R_i, \boldsymbol{\theta}_k) \\ = & \; \frac{P(V_i, C_i, S_i, N_i, L_i, R_i|\boldsymbol{\theta}_k)}{\int P(V_i, C_i, S_i, N_i, L_i, R_i|\boldsymbol{\theta}_k) d_{L_i}}. \end{aligned}$$

By Equation (A5), we can further write $P(L_i|O_i, \boldsymbol{\theta}_k)$ as

$$\begin{aligned} &P(L_i|O_i,\boldsymbol{\theta}_k) \\ &= \frac{\frac{P(V_i|R_i,\boldsymbol{\theta}_k)P(C_i|R_i,\boldsymbol{\theta}_k)P(S_i,L_i|R_i,\boldsymbol{\theta}_k)P(N_i,L_i|R_i,\boldsymbol{\theta}_k)P(R_i|\boldsymbol{\theta}_k)}{P(L_i|R_i,\boldsymbol{\theta}_k)}}{\int \frac{P(V_i|R_i,\boldsymbol{\theta}_k)P(C_i|R_i,\boldsymbol{\theta}_k)P(S_i,L_i|R_i,\boldsymbol{\theta}_k)P(N_i,L_i|R_i,\boldsymbol{\theta}_k)P(R_i|\boldsymbol{\theta}_k)}{P(L_i|R_i,\boldsymbol{\theta}_k)} d_{L_i}} \\ &= \frac{P(S_i,L_i|R_i,\boldsymbol{\theta}_k)P(N_i,L_i|R_i,\boldsymbol{\theta}_k)}{P(L_i|R_i,\boldsymbol{\theta}_k)\int \frac{P(S_i,L_i|R_i,\boldsymbol{\theta}_k)P(N_i,L_i|R_i,\boldsymbol{\theta}_k)}{P(L_i|R_i,\boldsymbol{\theta}_k)} d_{L_i}}. \end{aligned} \tag{B20}$$

Substituting $P(S_i,L_i|R_i,\boldsymbol{\theta}_k)$, $P(N_i,L_i|R_i,\boldsymbol{\theta}_k)$, and $P(L_i|R_i,\boldsymbol{\theta}_k)$ in Equation (B20) with Equations (15), (16), and (17) respectively, we obtain

$$P(L_i|O_i,\boldsymbol{\theta}_k) = \begin{cases} Z_i^1 & \text{for } Y_i < L_i \\ Z_i^2 & \text{for } 0 < L_i < y_i \\ Z_i^3 & \text{for } N_i < L_i < S_i \\ Z_i^4 & \text{for } S_i < L_i < N_i \end{cases} \tag{B21}$$

where $Z_i^1 = \bar{\lambda}_L'^{R_i} e^{-\bar{\lambda}_L'^{R_i} L_i + \bar{\lambda}_L'^{R_i} Y_i}$, $Z_i^2 = \frac{\left(\bar{\lambda}_S^{R_i} - \bar{\lambda}_S'^{R_i} + \bar{\lambda}_N^{R_i} + \bar{\lambda}_L^{R_i} - \bar{\lambda}_N'^{R_i}\right) e^{-\left(\bar{\lambda}_S^{R_i} - \bar{\lambda}_S'^{R_i} + \bar{\lambda}_N^{R_i} + \bar{\lambda}_L^{R_i} - \bar{\lambda}_N'^{R_i}\right) L_i}}{1 - e^{-\left(\bar{\lambda}_S^{R_i} - \bar{\lambda}_S'^{R_i} + \bar{\lambda}_N^{R_i} + \bar{\lambda}_L^{R_i} - \bar{\lambda}_N'^{R_i}\right) y_i}}$, $Z_i^3 = \frac{\left(\bar{\lambda}_S^{R_i} + \bar{\lambda}_L^{R_i} - \bar{\lambda}_S'^{R_i}\right) e^{-\left(\bar{\lambda}_S^{R_i} + \bar{\lambda}_L^{R_i} - \bar{\lambda}_S'^{R_i}\right) L_i}}{e^{-\left(\bar{\lambda}_S^{R_i} + \bar{\lambda}_L^{R_i} - \bar{\lambda}_S'^{R_i}\right) N_i} - e^{-\left(\bar{\lambda}_S^{R_i} + \bar{\lambda}_L^{R_i} - \bar{\lambda}_S'^{R_i}\right) S_i}}$, $Z_i^4 = \frac{\left(\bar{\lambda}_N^{R_i} + \bar{\lambda}_L^{R_i} - \bar{\lambda}_N'^{R_i}\right) e^{-\left(\bar{\lambda}_N^{R_i} + \bar{\lambda}_L^{R_i} - \bar{\lambda}_N'^{R_i}\right) L_i}}{e^{-\left(\bar{\lambda}_N^{R_i} + \bar{\lambda}_L^{R_i} - \bar{\lambda}_N'^{R_i}\right) S_i} - e^{-\left(\bar{\lambda}_N^{R_i} + \bar{\lambda}_L^{R_i} - \bar{\lambda}_N'^{R_i}\right) N_i}}$, $Y_i = \max(S_i, N_i)$, and $y_i = \min(S_i, N_i)$.

Going back to Equation (B19), we can rewrite it as

$$\begin{aligned} Q(\boldsymbol{\theta}|\boldsymbol{\theta}_k) = &\sum_{i=1}^{M} \ln[P(R_i|\boldsymbol{\theta})] + \sum_{i=1}^{M} \ln[P(V_i|R_i,\boldsymbol{\theta})] + \sum_{i=1}^{M} \ln[P(C_i|R_i,\boldsymbol{\theta})] \\ &+ \sum_{i=1}^{M} \underbrace{\int \ln[P(S_i,L_i|R_i,\boldsymbol{\theta})]\, P(L_i|O_i,\boldsymbol{\theta}_k) d_{L_i}}_{F1} \\ &+ \sum_{i=1}^{M} \underbrace{\int \ln[P(N_i,L_i|R_i,\boldsymbol{\theta})]\, P(L_i|O_i,\boldsymbol{\theta}_k) d_{L_i}}_{F2} \\ &- \sum_{i=1}^{M} \underbrace{\int \ln[P(L_i|R_i,\boldsymbol{\theta})]\, P(L_i|O_i,\boldsymbol{\theta}_k) d_{L_i}}_{F3}. \end{aligned} \tag{B22}$$

Let us calculate $F1$, $F2$, and $F3$ in Equation (B22) first. Let indicator $I_i = 1$ if $S_i > N_i$ and $I_i = 0$ otherwise. Substituting $P(L_i|O_i,\boldsymbol{\theta}_k)$ with Equation (B21), we have

$$F1 = \int_{Y_i}^{\infty} Z_i^1[\ln(\lambda_S^{R_i}) + \ln(\lambda_L^{\prime R_i}) - \lambda_L^{\prime R_i}L_i - (\lambda_S^{R_i} + \lambda_L^{R_i} - \lambda_L^{\prime R_i})S_i]d_{L_i} + \int_0^{y_i} Z_i^2[\ln(\lambda_L^{R_i}) + \ln(\lambda_S^{\prime R_i}) - \lambda_S^{\prime R_i}S_i - (\lambda_S^{R_i} + \lambda_L^{R_i} - \lambda_S^{\prime R_i})L_i]d_{L_i} + I_i\int_{N_i}^{S_i} Z_i^3[\ln(\lambda_L^{R_i}) + \ln(\lambda_S^{\prime R_i}) - \lambda_S^{\prime R_i}S_i - (\lambda_S^{R_i} + \lambda_L^{R_i} - \lambda_S^{\prime R_i})L_i]d_{L_i} + (1 - I_i)\int_{S_i}^{N_i} Z_i^4[\ln(\lambda_S^{R_i}) + \ln(\lambda_L^{\prime R_i}) - \lambda_L^{\prime R_i}L_i - (\lambda_S^{R_i} + \lambda_L^{R_i} - \lambda_L^{\prime R_i})S_i]d_{L_i} \tag{B28}$$

$$F2 = \int_{Y_i}^{\infty} Z_i^1[\ln(\lambda_N^{R_i}) + \ln(\lambda_L^{\prime R_i}) - \lambda_L^{\prime R_i}L_i - (\lambda_N^{R_i} + \lambda_L^{R_i} - \lambda_L^{\prime R_i})N_i]d_{L_i} + \int_0^{y_i} Z_i^2[\ln(\lambda_L^{R_i}) + \ln(\lambda_N^{\prime R_i}) - \lambda_N^{\prime R_i}N_i - (\lambda_N^{R_i} + \lambda_L^{R_i} - \lambda_N^{\prime R_i})L_i]d_{L_i} + I_i\int_{N_i}^{S_i} Z_i^3[\ln(\lambda_N^{R_i}) + \ln(\lambda_L^{\prime R_i}) - \lambda_L^{\prime R_i}L_i - (\lambda_N^{R_i} + \lambda_L^{R_i} - \lambda_L^{\prime R_i})N_i]d_{L_i} + (1 - I_i)\int_{S_i}^{N_i} Z_i^4[\ln(\lambda_L^{R_i}) + \ln(\lambda_N^{\prime R_i}) - \lambda_N^{\prime R_i}N_i - (\lambda_N^{R_i} + \lambda_L^{R_i} - \lambda_N^{\prime R_i})L_i]d_{L_i} \tag{B29}$$

$$F3 = \int_{Y_i}^{\infty} Z_i^1[\ln(\lambda_L^{\prime R_i}) - \lambda_L^{\prime R_i}L_i]d_{L_i} + \int_0^{y_i} Z_i^2[\ln(\lambda_L^{R_i}) - \lambda_L^{R_i}L_i]d_{L_i} + I_i\int_{N_i}^{S_i} Z_i^3[\ln(\lambda_L^{\prime R_i}) - \lambda_L^{\prime R_i}L_i]d_{L_i} + (1 - I_i)\int_{S_i}^{N_i} Z_i^4[\ln(\lambda_L^{\prime R_i}) - \lambda_L^{\prime R_i}L_i]d_{L_i} \tag{B30}$$

Substituting $F1$, $F2$, and $F3$ in Equation (B22) with Equations (B28) to (B30) respectively and differentiating $Q(\boldsymbol{\theta}|\boldsymbol{\theta}_k)$ with respect to each parameter in $\boldsymbol{\theta}$, we have the following first-order partial derivatives.

$$\frac{\partial Q(\boldsymbol{\theta}|\boldsymbol{\theta}_k)}{\partial \lambda_S^{R_i}} = \sum_i [\int_{Y_i}^{\infty} Z_i^1(\frac{1}{\lambda_S^{R_i}} - S_i)d_{L_i} + \int_0^{y_i} Z_i^2(-L_i)d_{L_i} + I_i\int_{N_i}^{S_i} Z_i^3(-L_i)d_{L_i} + (1 - I_i)\int_{S_i}^{N_i} Z_i^4(\frac{1}{\lambda_S^{R_i}} - S_i)d_{L_i}] \tag{B31}$$

$$\frac{\partial Q(\boldsymbol{\theta}|\boldsymbol{\theta}_k)}{\partial \lambda_S^{\prime R_i}} = \sum_i [\int_0^{y_i} Z_i^2(\frac{1}{\lambda_S^{\prime R_i}} - S_i + L_i)d_{L_i} + I_i\int_{N_i}^{S_i} Z_i^3(\frac{1}{\lambda_S^{\prime R_i}} - S_i + L_i)d_{L_i}] \tag{B32}$$

$$\frac{\partial Q(\boldsymbol{\theta}|\boldsymbol{\theta}_k)}{\partial \lambda_N^{R_i}} = \sum_i [\int_{Y_i}^{\infty} Z_i^1(\frac{1}{\lambda_N^{R_i}} - N_i)d_{L_i} + \int_0^{y_i} Z_i^2(-L_i)d_{L_i} + I_i\int_{N_i}^{S_i} Z_i^3(\frac{1}{\lambda_N^{R_i}} - N_i)d_{L_i} + (1 - I_i)\int_{S_i}^{N_i} Z_i^4(-L_i)d_{L_i}] \tag{B33}$$

$$\frac{\partial Q(\boldsymbol{\theta}|\boldsymbol{\theta}_k)}{\partial \lambda_N^{\prime R_i}} = \sum_i [\int_0^{y_i} Z_i^2 (\frac{1}{\lambda_N^{\prime R_i}} - N_i + L_i) d_{L_i} + (1 - I_i) \int_{S_i}^{N_i} Z_i^4 (\frac{1}{\lambda_N^{\prime R_i}} - N_i + L_i) d_{L_i}] \tag{B34}$$

$$\frac{\partial Q(\boldsymbol{\theta}|\boldsymbol{\theta}_k)}{\partial \lambda_L^{R_i}} = \sum_i [\int_{Y_i}^{\infty} Z_i^1 (-S_i - N_i) d_{L_i} + \int_0^{y_i} Z_i^2 (\frac{1}{\lambda_L^{R_i}} - L_i) d_{L_i} + I_i \int_{N_i}^{S_i} Z_i^3 (\frac{1}{\lambda_L^{R_i}} - L_i - N_i) d_{L_i} + (1 - I_i) \int_{S_i}^{N_i} Z_i^4 (\frac{1}{\lambda_L^{R_i}} - S_i - L_i) d_{L_i}] \tag{B35}$$

$$\frac{\partial Q(\boldsymbol{\theta}|\boldsymbol{\theta}_k)}{\partial \lambda_L^{\prime R_i}} = \sum_i [\int_{Y_i}^{\infty} Z_i^1 (\frac{1}{\lambda_L^{\prime R_i}} - L_i + S_i + N_i) d_{L_i} + I_i \int_{N_i}^{S_i} Z_i^3 N_i d_{L_i} + (1 - I_i) \int_{S_i}^{N_i} Z_i^4 S_i d_{L_i}] \tag{B36}$$

$$\frac{\partial Q(\boldsymbol{\theta}|\boldsymbol{\theta}_k)}{\partial \lambda_V^{R_i}} = \sum_i [\frac{1}{\lambda_V^{R_i}} - V_i] \tag{B37}$$

$$\frac{\partial Q(\boldsymbol{\theta}|\boldsymbol{\theta}_k)}{\partial \lambda_C^{R_i}} = \sum_i [\frac{1}{\lambda_C^{R_i}} - C_i] \tag{B38}$$

$$\frac{\partial Q(\boldsymbol{\theta}|\boldsymbol{\theta}_k)}{\partial p_1} = \sum_i [\frac{R_i}{p_1} - \frac{(1 - R_i)}{(1 - p_1)}] \tag{B39}$$

Making the above-derived partial derivatives equal zero for both $R_i = 0$ and $R_i = 1$, we obtain the following equations.

$$\sum_i (1 - R_i) [\int_{Y_i}^{\infty} Z_i^1 (\frac{1}{\lambda_S^0} - S_i) d_{L_i} + \int_0^{y_i} Z_i^2 (-L_i) d_{L_i} + I_i \int_{N_i}^{S_i} Z_i^3 (-L_i) d_{L_i} + (1 - I_i) \int_{S_i}^{N_i} Z_i^4 (\frac{1}{\lambda_S^0} - S_i) d_{L_i}] = 0 \tag{B40}$$

$$\sum_i (1 - R_i) [\int_0^{y_i} Z_i^2 (\frac{1}{\lambda_S^{\prime 0}} - S_i + L_i) d_{L_i} + I_i \int_{N_i}^{S_i} Z_i^3 (\frac{1}{\lambda_S^{\prime 0}} - S_i + L_i) d_{L_i}] = 0 \tag{B41}$$

$$\sum_i (1 - R_i) [\int_{Y_i}^{\infty} Z_i^1 (\frac{1}{\lambda_N^0} - N_i) d_{L_i} + \int_0^{y_i} Z_i^2 (-L_i) d_{L_i} + I_i \int_{N_i}^{S_i} Z_i^3 (\frac{1}{\lambda_N^0} - N_i) d_{L_i} + (1 - I_i) \int_{S_i}^{N_i} Z_i^4 (-L_i) d_{L_i}] = 0 \tag{B42}$$

$$\sum_i (1 - R_i) [\int_0^{y_i} Z_i^2 (\frac{1}{\lambda_N^{\prime 0}} - N_i + L_i) d_{L_i} + (1 - I_i) \int_{S_i}^{N_i} Z_i^4 (\frac{1}{\lambda_N^{\prime 0}} - N_i + L_i) d_{L_i}] = 0 \tag{B43}$$

$$\sum_i (1 - R_i) [\int_{Y_i}^{\infty} Z_i^1 (-S_i - N_i) d_{L_i} + \int_0^{y_i} Z_i^2 (\frac{1}{\lambda_L^0} - L_i) d_{L_i} + I_i \int_{N_i}^{S_i} Z_i^3 (\frac{1}{\lambda_L^0} - L_i - N_i) d_{L_i} + (1 - I_i) \int_{S_i}^{N_i} Z_i^4 (\frac{1}{\lambda_L^0} - S_i - L_i) d_{L_i}] = 0 \tag{B44}$$

$$\sum_i (1-R_i)[\int_{Y_i}^{\infty} Z_i^1(\frac{1}{\lambda_L'^0} - L_i + S_i + N_i)d_{L_i} + I_i \int_{N_i}^{S_i} Z_i^3 N_i d_{L_i} + (1-I_i)\int_{S_i}^{N_i} Z_i^4 S_i d_{L_i}] = 0 \quad \text{(B45)}$$

$$\sum_i (1-R_i)[\frac{1}{\lambda_V^0} - V_i] = 0 \quad \text{(B46)}$$

$$\sum_i (1-R_i)[\frac{1}{\lambda_C^0} - C_i] = 0 \quad \text{(B47)}$$

$$\sum_i R_i[\int_{Y_i}^{\infty} Z_i^1(\frac{1}{\lambda_S^1} - S_i)d_{L_i} + \int_0^{y_i} Z_i^2(-L_i)d_{L_i} + I_i \int_{N_i}^{S_i} Z_i^3(-L_i)d_{L_i} + (1 - I_i)\int_{S_i}^{N_i} Z_i^4(\frac{1}{\lambda_S^1} - S_i)d_{L_i}] = 0 \quad \text{(B48)}$$

$$\sum_i R_i[\int_0^{y_i} Z_i^2(\frac{1}{\lambda_S'^1} - S_i + L_i)d_{L_i} + I_i \int_{N_i}^{S_i} Z_i^3(\frac{1}{\lambda_S'^1} - S_i + L_i)d_{L_i}] = 0 \quad \text{(B49)}$$

$$\sum_i R_i[\int_{Y_i}^{\infty} Z_i^1(\frac{1}{\lambda_N^1} - N_i)d_{L_i} + \int_0^{y_i} Z_i^2(-L_i)d_{L_i} + I_i \int_{N_i}^{S_i} Z_i^3(\frac{1}{\lambda_N^1} - N_i)d_{L_i} + (1 - I_i)\int_{S_i}^{N_i} Z_i^4(-L_i)d_{L_i}] = 0 \quad \text{(B50)}$$

$$\sum_i R_i[\int_0^{y_i} Z_i^2(\frac{1}{\lambda_N'^1} - N_i + L_i)d_{L_i} + (1-I_i)\int_{S_i}^{N_i} Z_i^4(\frac{1}{\lambda_N'^1} - N_i + L_i)d_{L_i}] = 0 \quad \text{(B51)}$$

$$\sum_i R_i[\int_{Y_i}^{\infty} Z_i^1(-S_i - N_i)d_{L_i} + \int_0^{y_i} Z_i^2(\frac{1}{\lambda_L^1} - L_i)d_{L_i} + I_i \int_{N_i}^{S_i} Z_i^3(\frac{1}{\lambda_L^1} - L_i - N_i)d_{L_i} + (1 - I_i)\int_{S_i}^{N_i} Z_i^4(\frac{1}{\lambda_L^1} - S_i - L_i)d_{L_i}] = 0 \quad \text{(B52)}$$

$$\sum_i R_i[\int_{Y_i}^{\infty} Z_i^1(\frac{1}{\lambda_L'^1} - L_i + S_i + N_i)d_{L_i} + I_i \int_{N_i}^{S_i} Z_i^3 N_i d_{L_i} + (1-I_i)\int_{S_i}^{N_i} Z_i^4 S_i d_{L_i}] = 0 \quad \text{(B53)}$$

$$\sum_i R_i[\frac{1}{\lambda_V^1} - V_i] = 0 \quad \text{(B54)}$$

$$\sum_i (1-R_i)[\frac{1}{\lambda_C^1} - C_i] = 0 \quad \text{(B55)}$$

$$\sum_i [\frac{R_i}{P_1} - \frac{(1-R_i)}{(1-P_1)}] = 0 \quad \text{(B56)}$$

Letting $\Gamma_i^1 = \int_{Y_i}^{\infty} Z_i^1 L_i d_{L_i}$, $\Gamma_i^2 = \int_0^{y_i} Z_i^2 L_i d_{L_i}$, $\Gamma_i^3 = \int_{N_i}^{S_i} Z_i^3 L_i d_{L_i}$ and $\Gamma_i^4 = \int_{S_i}^{N_i} Z_i^4 L_i d_{L_i}$ and solving Equations (B40) to (B56), we can obtain a closed-form optimal solution of $\boldsymbol{\theta}$ shown in the theorem.

In order to show that the obtained solution is a single optimal solution, we need to show the global concavity of function $Q(\boldsymbol{\theta}|\boldsymbol{\theta}_k)$ (as defined in Equation (12)), which is equivalent to show that the Hessian matrix of $Q(\boldsymbol{\theta}|\boldsymbol{\theta}_k)$ is negative definite (Boyd and Vandenberghe 2004). By observing first-order partial derivatives in Equations (B31) to (B39), we know that

$$\frac{\partial Q(\boldsymbol{\theta}|\boldsymbol{\theta}_k)}{\partial x_1 \partial x_2} = 0 \tag{B57}$$

for any two parameters $x_1, x_2 \in \boldsymbol{\theta}$ and $x_1 \neq x_2$. Hence, all off-diagonal terms of the Hessian matrix of $Q(\boldsymbol{\theta}|\boldsymbol{\theta}_k)$ equal zero. The main diagonal terms of the Hessian matrix are given below:

$$\frac{\partial^2 Q(\boldsymbol{\theta}|\boldsymbol{\theta}_k)}{\partial \lambda_S^{R_i} \partial \lambda_S^{R_i}} = \sum_i [-\frac{1}{\left(\lambda_S^{R_i}\right)^2}](\int_{Y_i}^{\infty} Z_i^1 d_{L_i} + (1 - I_i) \int_{S_i}^{N_i} Z_i^4 d_{L_i}) \tag{B58}$$

$$\frac{\partial^2 Q(\boldsymbol{\theta}|\boldsymbol{\theta}_k)}{\partial \lambda_S^{\prime R_i} \partial \lambda_S^{\prime R_i}} = \sum_i [-\frac{1}{\left(\lambda_S^{\prime R_i}\right)^2}](\int_0^{y_i} Z_i^2 d_{L_i} + I_i \int_{N_i}^{S_i} Z_i^3 d_{L_i}) \tag{B59}$$

$$\frac{\partial^2 Q(\boldsymbol{\theta}|\boldsymbol{\theta}_k)}{\partial \lambda_N^{R_i} \partial \lambda_N^{R_i}} = \sum_i [-\frac{1}{\left(\lambda_N^{R_i}\right)^2}](\int_{Y_i}^{\infty} Z_i^1 d_{L_i} + I_i \int_{N_i}^{S_i} Z_i^3 d_{L_i}) \tag{B60}$$

$$\frac{\partial^2 Q(\boldsymbol{\theta}|\boldsymbol{\theta}_k)}{\partial \lambda_N^{\prime R_i} \partial \lambda_N^{\prime R_i}} = \sum_i [-\frac{1}{\left(\lambda_N^{\prime R_i}\right)^2}](\int_0^{y_i} Z_i^2 d_{L_i} + (1 - I_i) \int_{S_i}^{N_i} Z_i^4 d_{L_i}) \tag{B61}$$

$$\frac{\partial^2 Q(\boldsymbol{\theta}|\boldsymbol{\theta}_k)}{\partial \lambda_L^{R_i} \partial \lambda_L^{R_i}} = \sum_i [-\frac{1}{\left(\lambda_L^{R_i}\right)^2}](\int_0^{y_i} Z_i^2 d_{L_i} + I_i \int_{N_i}^{S_i} Z_i^3 d_{L_i} + (1 - I_i) \int_{S_i}^{N_i} Z_i^4 d_{L_i}) \tag{B62}$$

$$\frac{\partial^2 Q(\boldsymbol{\theta}|\boldsymbol{\theta}_k)}{\partial \lambda_L^{\prime R_i} \partial \lambda_L^{\prime R_i}} = \sum_i [-\frac{1}{\left(\lambda_L^{\prime R_i}\right)^2}](\int_{Y_i}^{\infty} Z_i^1 d_{L_i}) \tag{B63}$$

$$\frac{\partial^2 Q(\boldsymbol{\theta}|\boldsymbol{\theta}_k)}{\partial \lambda_V^{R_i} \partial \lambda_V^{R_i}} = \sum_i [-\frac{1}{\left(\lambda_V^{R_i}\right)^2}] \tag{B64}$$

$$\frac{\partial^2 Q(\boldsymbol{\theta}|\boldsymbol{\theta}_k)}{\partial \lambda_C^{R_i} \partial \lambda_C^{R_i}} = \sum_i [-\frac{1}{\left(\lambda_C^{R_i}\right)^2}] \tag{B64}$$

$$\frac{\partial^2 Q(\boldsymbol{\theta}|\boldsymbol{\theta}_k)}{\partial p_1 \partial p_1} = \sum_i [-\frac{R_i}{(p_1)^2} - \frac{(1 - R_i)}{(1 - p_1)^2}] \tag{B65}$$

It is easy to tell that all main diagonal terms of the Hessian matrix, shown in Equations (B58) to (B65), are negative. Therefore the Hessian matrix is negative definite (Dowling 1980). [QED]

**Appendix C. Estimation of $\boldsymbol{\theta}_0$**

Given a random sample $O_S$ of the training data with $M'$ records, i.e., $O_S = \{O_i\}$, where $O_i = < V_i, C_i, S_i, N_i, R_i >$ and $i = 1,2,\ldots,M'$, parameters in $\boldsymbol{\theta}_0$ is estimated as:

$$p_1 = \frac{\sum_{i=1}^{M'} R_i}{M'}, \tag{C1}$$

$$p_0 = 1 - p_1, \tag{C2}$$

$$\lambda_V^1 = \frac{\sum_{i=1}^{M'} R_i}{\sum_{i=1}^{M'} R_i V_i}, \tag{C3}$$

$$\lambda_V^0 = \frac{\sum_{i=1}^{M'} (1 - R_i)}{\sum_{i=1}^{M'} (1 - R_i) V_i}, \tag{C4}$$

$$\lambda_C^1 = \frac{\sum_{i=1}^{M'} R_i}{\sum_{i=1}^{M'} R_i C_i}, \tag{C5}$$

$$\lambda_C^0 = \frac{\sum_{i=1}^{M'} (1 - R_i)}{\sum_{i=1}^{M'} (1 - R_i) C_i}, \tag{C6}$$

$$\lambda_S^1 = \lambda_S'^1 = \frac{\sum_{i=1}^{M'} R_i}{\sum_{i=1}^{M'} R_i S_i}, \tag{C7}$$

$$\lambda_S^0 = \lambda_S'^0 = \frac{\sum_{i=1}^{M'} (1 - R_i)}{\sum_{i=1}^{M'} (1 - R_i) S_i}, \tag{C8}$$

$$\lambda_N^1 = \lambda_N'^1 = \frac{\sum_{i=1}^{M'} R_i}{\sum_{i=1}^{M'} R_i N_i}, \tag{C9}$$

$$\lambda_N^0 = \lambda_N'^0 = \frac{\sum_{i=1}^{M'} (1 - R_i)}{\sum_{i=1}^{M'} (1 - R_i) N_i}, \tag{C10}$$

$$\lambda_L^1 = \lambda_L'^1 = \frac{2 \cdot \sum_{i=1}^{M'} R_i}{\sum_{i=1}^{M'} R_i (S_i + N_i)}, \tag{C11}$$

$$\lambda_L^0 = \lambda_L'^0 = \frac{2 \cdot \sum_{i=1}^{M'} (1 - R_i)}{\sum_{i=1}^{M'} (1 - R_i)(S_i + N_i)}. \tag{C12}$$

We note that factor $L$ is unobserved, but depends on factors $S$ and $N$. We thus initialize the estimates for $L$ as the average of the estimates for $S$ and $N$. More concretely, $\frac{1}{\lambda_L^1} = \frac{1}{\lambda_L'^1} = (\frac{1}{\lambda_S^1} + \frac{1}{\lambda_N^1})/2$ and $\frac{1}{\lambda_L^0} = \frac{1}{\lambda_L'^0} = (\frac{1}{\lambda_S^0} + \frac{1}{\lambda_N^0})/2$.

**Appendix D. Computing Recommendation Probability**

We have

$$P\big(R_a = 1 \big| V_a, C_a, S_a, N_a, \widehat{\boldsymbol{\theta}}\big) = \frac{\int P\big(V_a, C_a, S_a, N_a, L_a, R_a = 1 \big| \widehat{\boldsymbol{\theta}}\big)\, d_{L_a}}{\sum_{r\in\{0,1\}} \int P\big(V_a, C_a, S_a, N_a, L_a, R_a = r \big| \widehat{\boldsymbol{\theta}}\big) d_{L_a}}. \tag{D1}$$

Let $INT(R_a) = \int P\big(V_a, C_a, S_a, N_a, L_a, R_a \big| \widehat{\boldsymbol{\theta}}\big) d_{L_a}$. We can rewrite Equation (D1) as

$$P(R_a = 1 | V_a, C_a, S_a, N_a, \widehat{\boldsymbol{\theta}}) = \frac{INT(R_a = 1)}{\sum_{r\in\{0,1\}} INT(R_a = r)} \tag{D2}$$

By Equation (A5), we have

$$\begin{aligned} INT(R_a) = & \int P\big(V_a, C_a, S_a, N_a, L_a, R_a \big| \widehat{\boldsymbol{\theta}}\big) d_{L_a} \\ = & \int \frac{P\big(V_a \big| R_a, \widehat{\boldsymbol{\theta}}\big) P\big(C_a | R_a, \widehat{\boldsymbol{\theta}}\big) P\big(S_a, L_a \big| R_a, \widehat{\boldsymbol{\theta}}\big) P\big(N_a, L_a \big| R_a, \widehat{\boldsymbol{\theta}}\big) P\big(R_a \big| \widehat{\boldsymbol{\theta}}\big)}{P\big(L_a \big| R_a, \widehat{\boldsymbol{\theta}}\big)} d_{L_a}. \end{aligned}$$

Given $\widehat{\boldsymbol{\theta}}$, each term in the right-hand side of the above equation can be calculated using Equations (15) to (17) as well as $P(V|R)$, $P(C|R)$, and $P(R)$ defined in §4.2 respectively. We therefore obtain

$$\begin{aligned} INT(R_a) = & \underbrace{\gamma(R_a) \cdot \int_{Y_a}^{\infty} \hat{\lambda}_S^{Ra} \hat{\lambda}_N^{Ra} \hat{\lambda}_L^{\prime Ra} e^{-\left(\hat{\lambda}_S^{Ra} + \hat{\lambda}_L^{Ra} - \hat{\lambda}_L^{\prime Ra}\right) S_a - \left(\hat{\lambda}_N^{Ra} + \hat{\lambda}_L^{Ra} - \hat{\lambda}_L^{\prime Ra}\right) N_a - \hat{\lambda}_L^{\prime Ra} L_a} d_{L_a}}_{H1(R_a)} \\ & \underbrace{+ \gamma(R_a) \cdot \int_0^{y_a} \hat{\lambda}_S^{\prime Ra} \hat{\lambda}_L^{Ra} \hat{\lambda}_N^{\prime Ra} e^{-\hat{\lambda}_S^{\prime Ra} S_a - \hat{\lambda}_N^{\prime Ra} N_a - \left(\hat{\lambda}_S^{Ra} - \hat{\lambda}_S^{\prime Ra} + \hat{\lambda}_N^{Ra} - \hat{\lambda}_N^{\prime Ra} + \hat{\lambda}_L^{Ra}\right) L_a} d_{L_a}}_{H2(R_a)} \\ & \underbrace{+ I_a \gamma(R_a) \cdot \int_{N_a}^{S_a} \hat{\lambda}_L^{Ra} \hat{\lambda}_S^{\prime Ra} \hat{\lambda}_N^{Ra} e^{-\hat{\lambda}_S^{\prime Ra} S_a - \left(\hat{\lambda}_N^{Ra} + \hat{\lambda}_L^{Ra} - \hat{\lambda}_L^{\prime Ra}\right) N_a - \left(\hat{\lambda}_S^{Ra} + \hat{\lambda}_L^{Ra} - \hat{\lambda}_S^{\prime Ra}\right) L_a} d_{L_a}}_{H3(R_a)} \\ & \underbrace{+ (1 - I_a) \gamma(R_a) \cdot \int_{S_a}^{N_a} \hat{\lambda}_L^{Ra} \hat{\lambda}_N^{\prime Ra} \hat{\lambda}_S^{Ra} e^{-\hat{\lambda}_N^{\prime Ra} N_a - \left(\hat{\lambda}_S^{Ra} + \hat{\lambda}_L^{Ra} - \hat{\lambda}_L^{\prime Ra}\right) S_a - \left(\hat{\lambda}_N^{Ra} + \hat{\lambda}_L^{Ra} - \hat{\lambda}_N^{\prime Ra}\right) L_a} d_{L_a}}_{H4(R_a)} \end{aligned}$$

where

$\gamma(R_a) = \hat{p}_{R_a} \hat{\lambda}_V^{Ra} \hat{\lambda}_C^{Ra} e^{-\left(\hat{\lambda}_V^{Ra} V_a + \hat{\lambda}_C^{Ra} C_a\right)}$,

$I_a = \begin{cases} 1 & \text{if } S_a > N_a, \\ 0 & \text{otherwise,} \end{cases}$

$Y_a = \max(S_a, N_a)$, and

$y_a = \min(S_a, N_a)$.

Solving $H1(R_a)$ to $H4(R_a)$, we have

$$H1(R_a) = \gamma(R_a) \cdot \hat{\lambda}_S^{Ra} \hat{\lambda}_N^{Ra} e^{-\left(\hat{\lambda}_S^{Ra} + \hat{\lambda}_L^{Ra} - \hat{\lambda}_L^{\prime Ra}\right) S_a - \left(\hat{\lambda}_N^{Ra} + \hat{\lambda}_L^{Ra} - \hat{\lambda}_L^{\prime Ra}\right) N_a - \hat{\lambda}_L^{\prime Ra} Y_a},$$

$$H2(R_a) = \gamma(R_a) \cdot \frac{\hat{\lambda}_S^{\prime Ra} \hat{\lambda}_L^{Ra} \hat{\lambda}_N^{\prime Ra} e^{-\hat{\lambda}_S^{\prime Ra} S_a - \hat{\lambda}_N^{\prime Ra} N_a} [1 - e^{-\left(\hat{\lambda}_S^{Ra} - \hat{\lambda}_S^{\prime Ra} + \hat{\lambda}_N^{Ra} - \hat{\lambda}_N^{\prime Ra} + \hat{\lambda}_L^{Ra}\right) y_a}]}{\hat{\lambda}_S^{Ra} - \hat{\lambda}_S^{\prime Ra} + \hat{\lambda}_N^{Ra} - \hat{\lambda}_N^{\prime Ra} + \hat{\lambda}_L^{Ra}},$$

$$H3(R_a) = I_a \gamma(R_a)$$

$$\cdot \frac{\hat{\lambda}_L^{R_a} \hat{\lambda}_S^{\prime R_a} \hat{\lambda}_N^{R_a} e^{-\hat{\lambda}_S^{\prime R_a} S_a - \left(\hat{\lambda}_N^{R_a} + \hat{\lambda}_L^{R_a} - \hat{\lambda}_L^{\prime R_a}\right) N_a} \left[ e^{-\left(\hat{\lambda}_S^{R_a} + \hat{\lambda}_L^{R_a} - \hat{\lambda}_S^{\prime R_a}\right) N_a} - e^{-\left(\hat{\lambda}_S^{R_a} + \hat{\lambda}_L^{R_a} - \hat{\lambda}_S^{\prime R_a}\right) S_a} \right]}{\hat{\lambda}_S^{R_a} + \hat{\lambda}_L^{R_a} - \hat{\lambda}_S^{\prime R_a}},$$

$$H4(R_a) = (1 - I_a) \gamma(R_a)$$

$$\cdot \frac{\hat{\lambda}_L^{R_a} \hat{\lambda}_N^{\prime R_a} \hat{\lambda}_S^{R_a} e^{-\hat{\lambda}_N^{\prime R_a} N_a - \left(\hat{\lambda}_S^{R_a} + \hat{\lambda}_L^{R_a} - \hat{\lambda}_L^{\prime R_a}\right) S_a} \left[ e^{-\left(\hat{\lambda}_N^{R_a} + \hat{\lambda}_L^{R_a} - \hat{\lambda}_N^{\prime R_a}\right) S_a} - e^{-\left(\hat{\lambda}_N^{R_a} + \hat{\lambda}_L^{R_a} - \hat{\lambda}_N^{\prime R_a}\right) N_a} \right]}{\hat{\lambda}_N^{R_a} + \hat{\lambda}_L^{R_a} - \hat{\lambda}_N^{\prime R_a}}.$$

Substituting $INT(R_a)$ in Equation (D2) with $H1(R_a) + H2(R_a) + H3(R_a) + H4(R_a)$, we obtain

$$P\left(R_a = 1 \middle| V_a, C_a, S_a, N_a, \widehat{\boldsymbol{\theta}}\right) = \frac{H1(R_a = 1) + H2(R_a = 1) + H3(R_a = 1) + H4(R_a = 1)}{\sum_{r \in \{0,1\}} [H1(R_a = r) + H2(R_a = r) + H3(R_a = r) + H4(R_a = r)]}.$$

**Appendix E. Robustness Analysis**

We conducted additional experiments by varying cost estimation or $K$. As shown in Tables E1-E8, our method substantially outperforms each benchmark method in every prediction month, under different cost estimation or $K$.

| **Prediction Month** ($t$+1) | **AA** | **CN** | **Jaccard** | **Katz** | **SVM** | **Our Method** |
|---|---|---|---|---|---|---|
| 3 | 0.18 | 0.17 | 0.05 | 0.25 | 0.28 | 0.47 |
| 4 | 0.19 | 0.20 | 0.04 | 0.27 | 0.31 | 0.36 |
| 5 | 0.16 | 0.16 | 0.05 | 0.20 | 0.26 | 0.35 |
| 6 | 0.16 | 0.20 | 0.03 | 0.23 | 0.27 | 0.42 |
| 7 | 0.17 | 0.16 | 0.03 | 0.22 | 0.26 | 0.39 |
| 8 | 0.18 | 0.18 | 0.03 | 0.25 | 0.25 | 0.39 |
| 9 | 0.16 | 0.17 | 0.03 | 0.22 | 0.25 | 0.36 |
| 10 | 0.16 | 0.16 | 0.02 | 0.22 | 0.24 | 0.37 |
| 11 | 0.19 | 0.20 | 0.02 | 0.27 | 0.29 | 0.36 |
| 12 | 0.20 | 0.21 | 0.03 | 0.28 | 0.30 | 0.39 |
| Mean | 0.17 | 0.18 | 0.03 | 0.24 | 0.27 | 0.39 |
| Std. | 0.01 | 0.02 | 0.01 | 0.03 | 0.02 | 0.04 |

**Table E1:** Top-$K$ Utility-based Precision: Our Method versus Benchmark Methods ($\rho = 0.5$)[10]

| **Prediction Month** ($t$+1) | **AA** | **CN** | **Jaccard** | **Katz** | **SVM** | **Our Method** |
|---|---|---|---|---|---|---|
| 3 | \$0.31 | \$0.24 | \$0.10 | \$0.47 | \$0.97 | \$1.16 |
| 4 | \$0.45 | \$0.49 | \$0.12 | \$0.73 | \$0.93 | \$1.14 |
| 5 | \$0.28 | \$0.27 | \$0.06 | \$0.34 | \$0.77 | \$1.02 |
| 6 | \$0.32 | \$0.35 | \$0.07 | \$0.39 | \$0.56 | \$0.84 |
| 7 | \$0.29 | \$0.28 | \$0.05 | \$0.41 | \$0.57 | \$0.79 |
| 8 | \$0.38 | \$0.39 | \$0.05 | \$0.56 | \$0.86 | \$0.97 |
| 9 | \$0.37 | \$0.38 | \$0.05 | \$0.53 | \$0.84 | \$1.00 |
| 10 | \$0.39 | \$0.40 | \$0.05 | \$0.57 | \$0.83 | \$1.18 |
| 11 | \$0.62 | \$0.66 | \$0.06 | \$1.00 | \$1.43 | \$1.78 |
| 12 | \$0.67 | \$0.73 | \$0.06 | \$1.09 | \$1.57 | \$1.88 |
| Mean | \$0.41 | \$0.42 | \$0.07 | \$0.61 | \$0.93 | \$1.18 |
| Std. | \$0.13 | \$0.16 | \$0.03 | \$0.26 | \$0.33 | \$0.37 |

**Table E2:** Average Utility: Our Method versus Benchmark Methods ($\rho = 0.5$)

[10] Top-$K$ utility-based precision of benchmark methods does not change with different cost estimations because these methods do not consider cost when recommending links.

| Prediction Month (*t*+1) | AA | CN | Jaccard | Katz | SVM | Our Method |
|---|---|---|---|---|---|---|
| 3 | 0.18 | 0.17 | 0.05 | 0.25 | 0.28 | 0.43 |
| 4 | 0.19 | 0.20 | 0.04 | 0.27 | 0.31 | 0.38 |
| 5 | 0.16 | 0.16 | 0.05 | 0.20 | 0.26 | 0.42 |
| 6 | 0.16 | 0.20 | 0.03 | 0.23 | 0.27 | 0.42 |
| 7 | 0.17 | 0.16 | 0.03 | 0.22 | 0.26 | 0.41 |
| 8 | 0.18 | 0.18 | 0.03 | 0.25 | 0.25 | 0.36 |
| 9 | 0.16 | 0.17 | 0.03 | 0.22 | 0.25 | 0.41 |
| 10 | 0.16 | 0.16 | 0.02 | 0.22 | 0.24 | 0.41 |
| 11 | 0.19 | 0.20 | 0.02 | 0.27 | 0.29 | 0.37 |
| 12 | 0.20 | 0.21 | 0.03 | 0.28 | 0.30 | 0.40 |
| Mean | 0.17 | 0.18 | 0.03 | 0.24 | 0.27 | 0.40 |
| Std. | 0.01 | 0.02 | 0.01 | 0.03 | 0.02 | 0.02 |

**Table E3:** Top-*K* Utility-based Precision: Our Method versus Benchmark Methods ($\rho = 2$)

| Prediction Month (*t*+1) | AA | CN | Jaccard | Katz | SVM | Our Method |
|---|---|---|---|---|---|---|
| 3 | $0.28 | $0.20 | $0.08 | $0.43 | $0.91 | $1.09 |
| 4 | $0.44 | $0.46 | $0.12 | $0.70 | $0.73 | $1.09 |
| 5 | $0.25 | $0.23 | $0.06 | $0.30 | $0.49 | $1.00 |
| 6 | $0.28 | $0.31 | $0.07 | $0.35 | $0.47 | $0.82 |
| 7 | $0.25 | $0.23 | $0.04 | $0.36 | $0.47 | $0.72 |
| 8 | $0.36 | $0.36 | $0.05 | $0.54 | $0.51 | $0.91 |
| 9 | $0.35 | $0.35 | $0.05 | $0.49 | $0.73 | $0.95 |
| 10 | $0.35 | $0.35 | $0.04 | $0.52 | $0.64 | $1.06 |
| 11 | $0.58 | $0.60 | $0.05 | $0.95 | $1.13 | $1.42 |
| 12 | $0.62 | $0.66 | $0.04 | $1.02 | $1.37 | $1.67 |
| Mean | $0.38 | $0.38 | $0.06 | $0.57 | $0.74 | $1.07 |
| Std. | $0.13 | $0.16 | $0.02 | $0.25 | $0.31 | $0.28 |

**Table E4:** Average Utility: Our Method versus Benchmark Methods ($\rho = 2$)

| Prediction Month (*t*+1) | AA | CN | Jaccard | Katz | SVM | Our Method |
|---|---|---|---|---|---|---|
| 3 | 0.09 | 0.12 | 0.03 | 0.23 | 0.29 | 0.43 |
| 4 | 0.13 | 0.16 | 0.02 | 0.27 | 0.31 | 0.43 |
| 5 | 0.12 | 0.13 | 0.05 | 0.20 | 0.24 | 0.42 |
| 6 | 0.15 | 0.16 | 0.02 | 0.24 | 0.25 | 0.36 |
| 7 | 0.12 | 0.16 | 0.02 | 0.26 | 0.29 | 0.32 |
| 8 | 0.13 | 0.16 | 0.02 | 0.24 | 0.26 | 0.36 |
| 9 | 0.12 | 0.15 | 0.01 | 0.21 | 0.27 | 0.35 |
| 10 | 0.12 | 0.15 | 0.01 | 0.21 | 0.30 | 0.33 |
| 11 | 0.14 | 0.17 | 0.01 | 0.25 | 0.31 | 0.43 |
| 12 | 0.15 | 0.19 | 0.01 | 0.27 | 0.30 | 0.40 |
| Mean | 0.13 | 0.16 | 0.02 | 0.24 | 0.28 | 0.38 |
| Std. | 0.02 | 0.02 | 0.01 | 0.02 | 0.03 | 0.04 |

**Table E5:** Top-*K* Utility-based Precision: Our Method versus Benchmark Methods (*K*=0.25%× number of potential links in a month)

| Prediction Month (*t*+1) | AA | CN | Jaccard | Katz | SVM | Our Method |
|---|---|---|---|---|---|---|
| 3 | $0.20 | $0.25 | $0.16 | $0.66 | $1.21 | $1.62 |
| 4 | $0.58 | $0.68 | $0.21 | $1.17 | $1.35 | $1.74 |
| 5 | $0.35 | $0.36 | $0.12 | $0.46 | $1.11 | $1.50 |
| 6 | $0.41 | $0.47 | $0.11 | $0.66 | $0.79 | $1.14 |
| 7 | $0.36 | $0.40 | $0.05 | $0.67 | $0.81 | $1.00 |
| 8 | $0.46 | $0.56 | $0.07 | $0.80 | $0.91 | $1.26 |
| 9 | $0.45 | $0.53 | $0.06 | $0.76 | $0.96 | $1.39 |
| 10 | $0.48 | $0.58 | $0.06 | $0.84 | $1.27 | $1.46 |
| 11 | $0.78 | $1.00 | $0.06 | $1.51 | $1.93 | $2.59 |
| 12 | $0.81 | $1.11 | $0.06 | $1.72 | $2.02 | $2.37 |
| Mean | $0.49 | $0.59 | $0.10 | $0.93 | $1.24 | $1.61 |
| Std. | $0.19 | $0.27 | $0.05 | $0.41 | $0.43 | $0.51 |

**Table E6:** Average Utility: Our Method versus Benchmark Methods (*K*=0.25%× number of potential links in a month)

| Prediction Month (*t*+1) | AA | CN | Jaccard | Katz | SVM | Our Method |
|---|---|---|---|---|---|---|
| 3 | 0.24 | 0.21 | 0.07 | 0.26 | 0.28 | 0.42 |
| 4 | 0.30 | 0.26 | 0.08 | 0.29 | 0.31 | 0.39 |
| 5 | 0.25 | 0.22 | 0.06 | 0.27 | 0.28 | 0.40 |
| 6 | 0.23 | 0.23 | 0.05 | 0.27 | 0.30 | 0.35 |
| 7 | 0.26 | 0.24 | 0.04 | 0.30 | 0.31 | 0.38 |
| 8 | 0.27 | 0.26 | 0.04 | 0.30 | 0.31 | 0.36 |
| 9 | 0.27 | 0.27 | 0.04 | 0.31 | 0.32 | 0.38 |
| 10 | 0.27 | 0.27 | 0.04 | 0.31 | 0.33 | 0.41 |
| 11 | 0.28 | 0.29 | 0.04 | 0.32 | 0.34 | 0.39 |
| 12 | 0.29 | 0.29 | 0.04 | 0.32 | 0.34 | 0.39 |
| Mean | 0.27 | 0.25 | 0.05 | 0.29 | 0.31 | 0.39 |
| Std. | 0.02 | 0.03 | 0.01 | 0.02 | 0.02 | 0.02 |

**Table E7:** Top-*K* Utility-based Precision: Our Method versus Benchmark Methods (*K*=0.75%× number of potential links in a month)

| Prediction Month (*t*+1) | AA | CN | Jaccard | Katz | SVM | Our Method |
|---|---|---|---|---|---|---|
| 3 | $0.40 | $0.27 | $0.07 | $0.41 | $0.69 | $0.84 |
| 4 | $0.47 | $0.47 | $0.09 | $0.59 | $0.66 | $0.76 |
| 5 | $0.29 | $0.27 | $0.05 | $0.34 | $0.56 | $0.76 |
| 6 | $0.30 | $0.31 | $0.06 | $0.41 | $0.53 | $0.63 |
| 7 | $0.32 | $0.29 | $0.04 | $0.42 | $0.46 | $0.56 |
| 8 | $0.41 | $0.41 | $0.04 | $0.53 | $0.55 | $0.70 |
| 9 | $0.43 | $0.44 | $0.04 | $0.54 | $0.56 | $0.77 |
| 10 | $0.44 | $0.44 | $0.04 | $0.57 | $0.67 | $0.76 |
| 11 | $0.65 | $0.67 | $0.06 | $0.84 | $0.89 | $1.13 |
| 12 | $0.70 | $0.72 | $0.05 | $0.90 | $0.93 | $1.22 |
| Mean | $0.44 | $0.43 | $0.05 | $0.56 | $0.65 | $0.81 |
| Std. | $0.14 | $0.16 | $0.02 | $0.19 | $0.16 | $0.21 |

**Table E8:** Average Utility: Our Method versus Benchmark Methods (*K*= *K*=0.75%× number of potential links in a month)

**Appendix F. Contribution of Each Component of Our Method to Its Performance**

Our method illustrated in Figure 3 consists of four components: the utility component including value (*V*) and cost (*C*), the structural proximity component (*S*), the nodal proximity component (*N*), and the latent linkage likelihood component (*L*). We conducted experiments to empirically evaluate the contribution of each component to the performance of our method, using the data and parameter values described in §5.1. In an experiment, we dropped one component (e.g., the utility component) from our method. The performance difference between our method with and without the dropped component indicates the contribution of the component to the performance of our method. As shown in Figures F1 and F2, dropping the utility component degrades the performance of our method by an average of 45.79% in terms of average utility and an average of 56.22% in terms of top-*K* utility-based precision. In addition, dropping the structural proximity, the nodal proximity, and the latent linkage component decreases the average utility of our method by an average of 15.29%, 15.20%, and 16.55% respectively and reduces the top-*K* utility-based precision of our method by an average of 44.69%, 44.38%, and 44.61% respectively. Our experimental results suggest that (i) each component contributes to the performance our method; and (ii) the utility component contributes the most, which echoes the importance of considering utility in link recommendation.

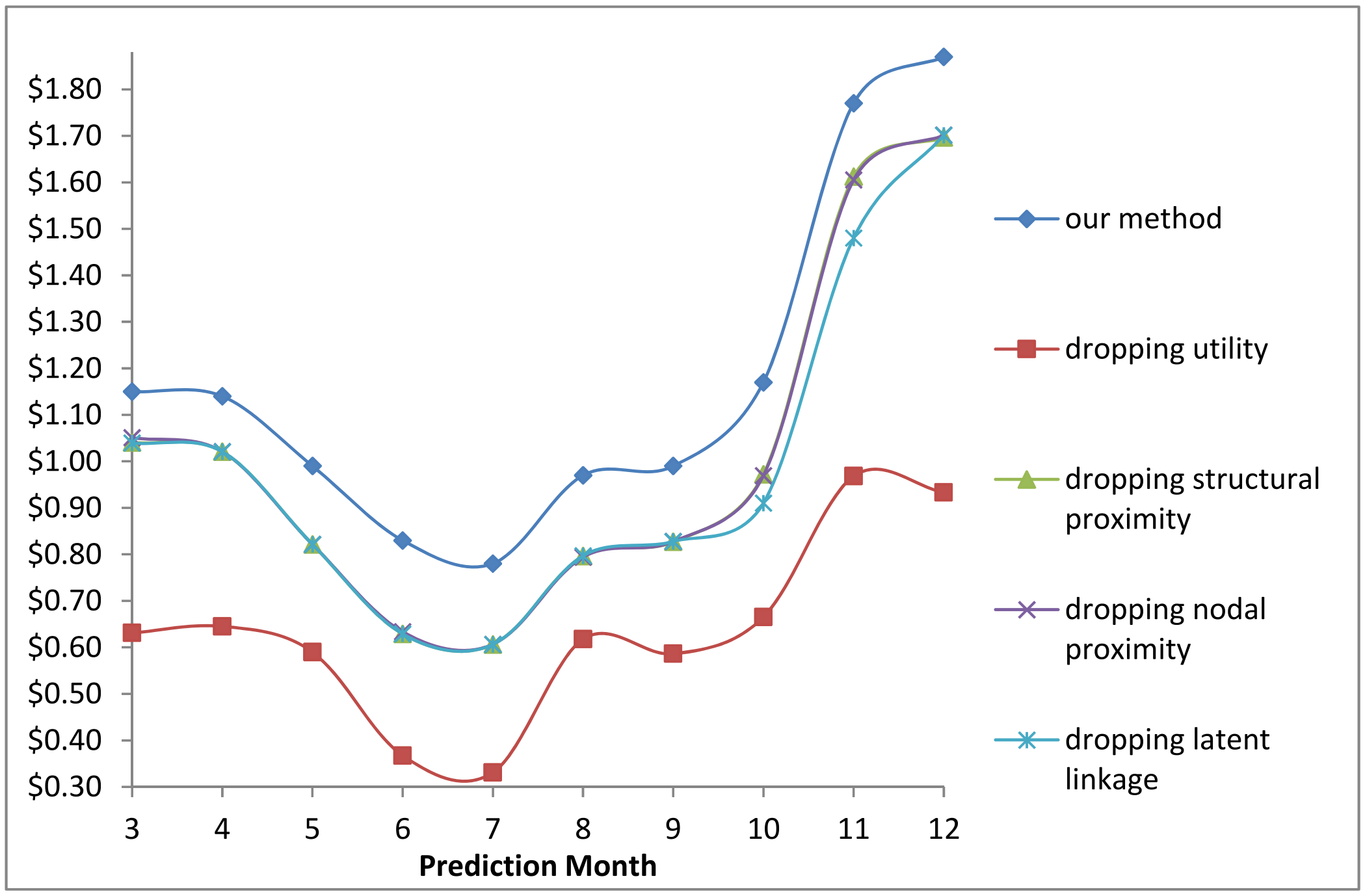

**Figure F1:** Average Utility: Our Method versus Our Method without a Component

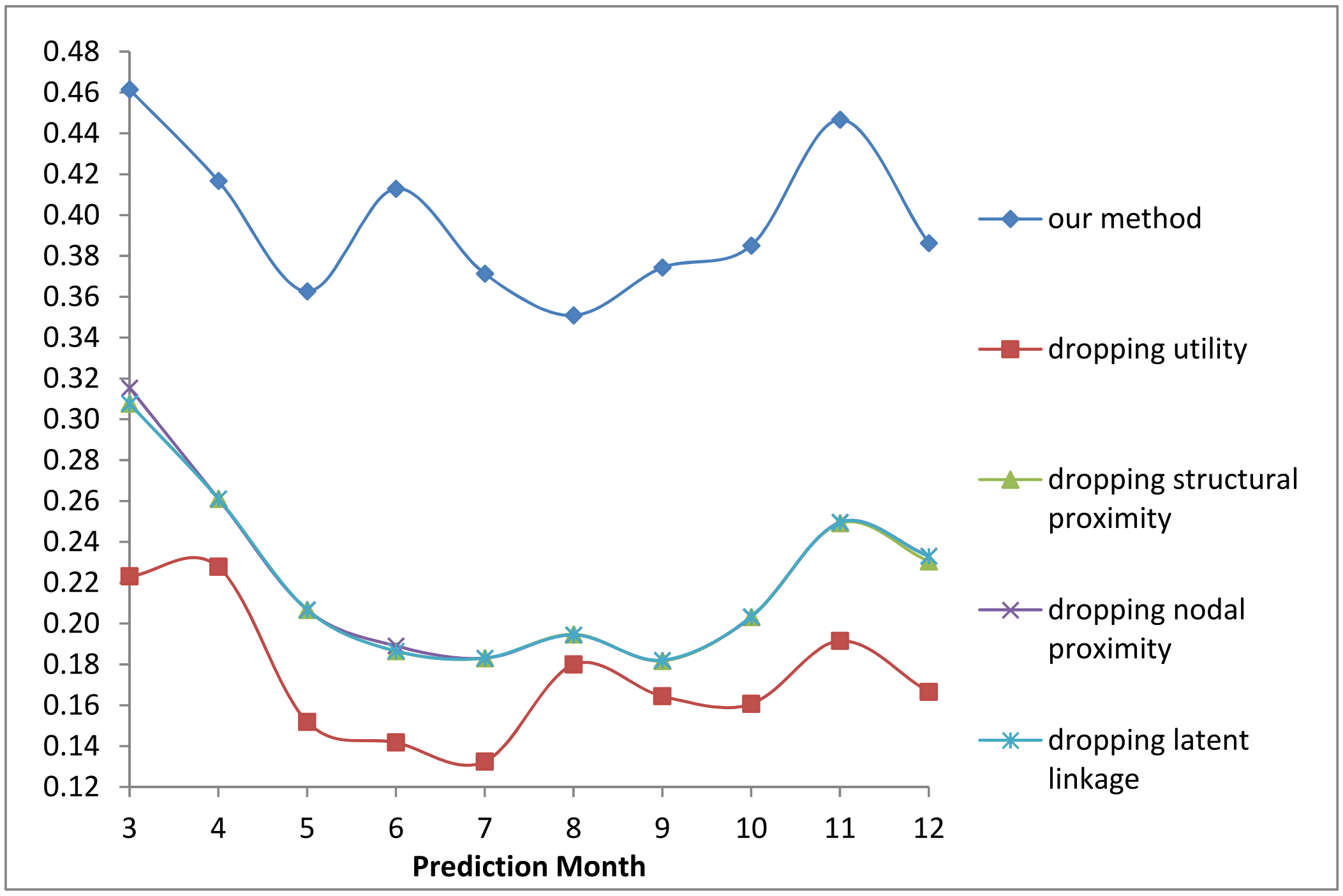

**Figure F2:** Top-*K* Utility-based Precision: Our Method versus Our Method without a Component

**Appendix G.** Performance Evaluation with another Social Network Data Set

We evaluated the performance of our method using a data set provided by another U.S. online social network. Our data span a three-month period in 2011 and contain information about who registered as a user of the online social network in which month as well as who became linked to whom in which month. As shown in Table G1, the number of users increases from 36,079 in month 1 to 341,023 in month 3 and the number of links grows from 77,098 in month 1 to 2,616,963 in month 3. Each user is described by a set of profile terms such as schools attended and places lived. This additional data set and the data set described in §5.1 were collected from different types of social networks. One is from a social network for connecting with friends and family members and the other one is from a social network for professional networking[11]. Thus, this additional data set allows us to evaluate the performance of our method in a different type of social network.

| Month | Accumulated Number of Users | Accumulated Number of Links |
|---|---|---|
| 1 | 36,079 | 77,098 |
| 2 | 198,095 | 283,732 |
| 3 | 341,023 | 2,616,963 |

**Table G1:** Accumulated Number of Users and Links

Using this additional data set, we conducted experiments to compare the performance between our method and benchmark methods listed in Table 1, by following the parameter calibration described in §5.1 and the experimental procedure illustrated in §5.2. Current month $t$ and prediction month $t + 1$ in our experiments are months 2 and 3 respectively[12]. In an experiment, our method as well as each benchmark method recommended top-*K* potential links out of potential links in month 2. Using data by prediction month 3, we verified whether a potential link in month 2 was actually accepted and established by users in month 3, computed its utility with Equation (6), and identified true top-*K* potential links that had the highest utilities. The performance of a method was then evaluated by comparing its recommended top-*K* potential links with true top-*K* potential links using metrics such as top-*K* utility-based precision and average utility. As discussed in §5.1, we focus on potential links that would connect users who are two hops away and the number of such potential links in month 2 is 42,608,775.

[11] Due to requests from data providers, we regretfully cannot disclose the identities of these two social networks.

[12] To construct training data for our method, we need data in month $t - 1$ and month $t$. Thus, current month $t$ in our experiments is month 2 instead of month 1.

Tables G2 to G4 compare the performance between our method and benchmark methods under different *K*. As shown, our method significantly outperforms each benchmark method across *K*, in terms of both top-*K* utility-based precision and average utility. Let us take SVM, the best performing benchmark method, as an example. Averaged across *K*, our method outperforms SVM by 24.27% in terms of top-*K* utility-based precision and 16.14% in terms of average utility. These experimental results provide additional evidence suggesting the performance advantage of our method over benchmark methods.

| | **AA** | **CN** | **Jaccard** | **Katz** | **SVM** | **Our Method** |
|---|---|---|---|---|---|---|
| Top-*K* Utility-based Precision | 0.10 | 0.12 | 0.003 | 0.21 | 0.25 | 0.31 |
| Average Utility | $1.14 | $1.31 | $0.05 | $1.98 | $2.54 | $2.94 |

**Table G2:**Our Method versus Benchmark Methods (*K*=0.25%× number of potential links in month 2)

| | **AA** | **CN** | **Jaccard** | **Katz** | **SVM** | **Our Method** |
|---|---|---|---|---|---|---|
| Top-*K* Utility-based Precision | 0.20 | 0.24 | 0.005 | 0.38 | 0.48 | 0.56 |
| Average Utility | $1.00 | $1.17 | $0.04 | $1.95 | $2.16 | $2.52 |

**Table G3:** Our Method versus Benchmark Methods (*K*=0.5%× number of potential links in month 2)

| | **AA** | **CN** | **Jaccard** | **Katz** | **SVM** | **Our Method** |
|---|---|---|---|---|---|---|
| Top-*K* Utility-based Precision | 0.27 | 0.31 | 0.009 | 0.47 | 0.52 | 0.68 |
| Average Utility | $0.99 | $1.11 | $0.04 | $1.79 | $1.99 | $2.30 |

**Table G4:**Our Method versus Benchmark Methods (*K*=0.75%× number of potential links in month 2)